\documentclass[pra,superscriptaddress,a4paper]{revtex4}

\usepackage[T1]{fontenc}
\usepackage{aas_macros}
\usepackage{textcomp}   
\usepackage{gensymb}
\usepackage[latin1]{inputenc}
\usepackage[english]{babel}
\usepackage[dvipsnames,pdftex]{xcolor}
\usepackage[pdftex]{graphicx,epsfig}
\usepackage{amsmath,amsthm,amssymb,amsfonts,amsbsy}
\usepackage[breaklinks=true,hidelinks,colorlinks=true,linkcolor = black,citecolor=black,urlcolor=blue]{hyperref}
\usepackage[caption=false]{subfig}
\usepackage{multirow}
\usepackage[tight,nice]{units}
\usepackage{siunitx}

\DeclareSIUnit{\persqrthz}{\ensuremath{/\sqrt{\text{\hertz}}}}

\newlength{\bildtitel}
\newcommand\REVIEW[1]{\message{LaTeX Warning: \noexpand untreated nEDM-REVIEW command in \jobname .tex: l\the\inputlineno}}
\newcommand{\mrm}{\mathrm}

\newcommand{\mbf}{\boldsymbol}
\renewcommand{\parallel}{\uparrow\!\uparrow}
\renewcommand{\nparallel}{\uparrow\!\downarrow}

\newcommand{\diff}[1]{\operatorname{d}\ifthenelse{\equal{#1}{}}{\,}{\!#1}}

\newcommand{\pow}[2]{\ensuremath{#1\!\times\!10^{#2}}}

\newcommand{\ecm}{\ensuremath{\si{\elementarycharge}\!\cdot\!\cm}}
\newcommand{\dn}{\ensuremath{d_\text{n}}}

\def\Bra#1{\left\langle#1\right|}
\def\Ket#1{\left|#1\right\rangle}

\newcommand{\stateup}{\ensuremath{\Ket{\uparrow}}}
\newcommand{\statedo}{\ensuremath{\Ket{\downarrow}}}

\newcommand{\muT}{\mbox{\micro T}}

\newcommand{\pT}{\mbox{pT}}

\newcommand{\cm}{\ensuremath{\mathrm{cm}}}

\newcommand{\tHe}{\ensuremath{{}^3\mathrm{He}}}

\newcommand{\magHg}{\ensuremath{{}^{199}\text{Hg}}}

\newcommand{\magXe}{\ensuremath{{}^{129}\mathrm{Xe}}}

\addto\captionsenglish{}
\addto\captionsenglish{}

\graphicspath{{./art/}}
\begin{document}

\title{The quest to find an electric dipole moment of the neutron}

\author{P. Schmidt-Wellenburg}
\email{philipp.schmidt-wellenburg@psi.ch} \affiliation{Paul Scherrer Institut, 5232 Villigen PSI,
Switzerland} 

\begin{abstract}
Until now no electric dipole moment of the neutron (nEDM) has been observed. Why it is so vanishingly small, escaping detection for the last 65 years, is not easy to explain. In general it is considered as one of the most sensitive probes for the violation of the combined symmetry of charge and parity (CP). A discovery could shed light on the poorly understood matter/antimatter asymmetry of the Universe. The neutron EDM might one day help to distinguish different sources of CP-violation in combination with measurements of paramagnetic molecules, diamagnetic atoms and other nuclei.
This review presents an overview of the most important concepts in searches for an nEDM as well as a brief overview of the worldwide efforts.
\end{abstract}

\maketitle

\section{INTRODUCTION}
One of the most intriguing questions in cosmology and perhaps even in all of physics is: ``Why is there so much matter in the Universe and so little antimatter?'' This observation is even more puzzling as the near-perfect symmetry of particles and anti-particles is a firmly established fact of collider physics.
Outside the laboratories of particle physics antimatter can be seen only in cosmic rays which are a result of particle collisions in the atmosphere consistent with the same processes observed in particle accelerators or in the decay of $^{40}$K, the only natural positron emitter with a branching ratio of \SI{12}{ppm}\,\cite{Engelkemeir1962PR}. 
 Until today there is no evidence for any primordial antimatter within our galaxy or even beyond.
There is no indication for any form of co-existence of matter and antimatter in clusters or galaxies within our Universe. 
If such regions would exist, one would expect a distinct source of annihilation gammas from any boundary between an antimatter and matter part of the Universe, which has not been observed. 
Hence, it is usually conclude that our visible Universe is made entirely of matter and is intrinsic matter non-symmetric.

   The natural hypothesis for an initial state prior to the Big Bang is a fully symmetric state indicating the asymmetry of matter and antimatter $\eta = 0$. To describe this asymmetry I use the standard notation:
	
	\begin{equation}
		\eta = \frac{n_{\rm b}-n_{\rm \bar{b}}}{n_{\gamma}},
	\label{eq:BAU}
	\end{equation}
\noindent where $n_{\rm b}$ and $n_{\rm \overline{b}}$ are the  number densities of baryons and antibaryons (most matter is made out of baryons which are themselves particles made out of quarks and gluons, e.g.\ protons and neutrons), and $n_{\gamma}$ is the number density of photons emitted during the primordial process of matter/anitmatter creation. 
Starting from a symmetric state $\eta = 0$, the Big Bang would have created enormous quantities of baryons and antibaryons, which, while the Universe expanded, would have found each other and annihilated, emitting plenty of photons until the baryon/antibaryon density became low enough for the process to stop.
Assuming such a scenario including the known symmetry violations of the Standard Model of particle physics (SM) results in $\eta \approx 10^{-18}$\cite{Riotto1999ARNPS,Morrissey2012NJP}, while the observed asymmetry is eight orders of magnitude larger when derived from the measurement of the microwave background of the universe $\eta=6.1^{+0.3}_{-0.2}\!\times\!10^{-10}$\,\cite{Dine2012}, or from the abundance of light elements produced in primordial nucleosynthesis with $5.1\!\times\!10^{-10}<\eta<6.7\!\times\!10^{-10}$\,\cite{Cyburt2008JCAP}. The very simple and efficient solution would be to accept an initial condition which is asymmetric. Such a sort of fine tuning is strongly disfavored as it contradicts any naturalness principle. Hence, whatever the answer to this outstanding problem might be, it relates to a fundamental structural asymmetry of physics and might have left a faint footprint on all fundamental particles and their interactions.  

With the discovery of the Higgs boson\,\cite{Aad2012PLB,Chatrchyan2012PLB} particle physics at particle accelerators celebrated its latest success, following an exceptional series of discoveries that conformed with the SM, which beautifully describes all particles and interactions known from laboratory experiments. 
The possibility that new physics could be at mass scales beyond the reach of collider experiments or coupling very weakly to known particles, in combination with astrophysical observations (e.g.\ dark matter, neutrino oscillations) which cannot be explained by the SM has nurtured vivid interest in high-precision physics at low energies\,\cite{Raidal2008} in recent years.
Some of these experiments look for the remaining faint footprints of structural asymmetries which might have led to the observed baryon asymmetry of the Universe (BAU). One such high-precision search for new physics is the quest to find an electric dipole moment (EDM) of the neutron. 

A very intuitive picture of an EDM consists of two point charges, one with charge $+e$, the other with $-e$, separated by a distance $\mbf{r}$ giving rise to $\mbf{d} = e\cdot \mbf{r}$.  In a more general classical limit for a continuous charge distribution $\rho(\mbf{r})$ the EDM can be described by

\begin{equation}
	\mbf{d}^{\rm cl} = \int{\rm d\mbf{r}\,\mbf{r}\rho(\mbf{r})};
\label{eq:dnClassic}
\end{equation}
\noindent where $\mbf{r}$ is integrated over the extended object. This expression will be redefined in the section on symmetries and EDMs (Sec.\,\ref{sec:Symmetries}) for point-like quantum objects. Thus, the common unit in which EDMs are measured is $\ecm$: equivalent to a separation between two opposite charged elementary point charges. 

Any particle which has a well defined non-degenerate ground state and an EDM ($d\neq0$) violates {\it parity (P)}-symmetry and the symmetry under {\it time}-reversal {\it(T)}. This also indicates a violation of the combined symmetry of {\it charge (C)} and {\it P}, under the assumption of Lorentz and CPT-invariance. Thus, the discovery of an EDM of the neutron would indicate that {\it CP}-symmetry is violated, and this in turn may be connected to the structural asymmetry required to create a matter dominated Universe. {\it CP}-violation is one of three necessary conditions for a baryon asymmetry evolving from an initial symmetric state. These three Sakharov criteria\,\cite{Sakharov1991}, namely i) baryon number violation, ii) {\it C} and {\it CP}-symmetry violation, and iii) departure from thermal equilibrium, were identified as the minimal set of conditions which any theory must fulfill to explain the observed BAU\@.  For a more detailed explanation of the role of symmetries in EDMs see Sec.\,\ref{sec:Symmetries}.

While in the recent past remarkable progress has been made in the search for an electron EDM using the enormous electric field inside a ThO molecule\,\cite{Baron2014Science}, and also for the diamagnetic \magHg-EDM\,\cite{Graner2016PRL}, that of the neutron has slowed down. The latest result for the neutron EDM, $|\dn|\!<\!\unit[\pow{3}{-26}]{\ecm}$\,\cite{Pendlebury2015PRD}, arises from a re-analysis of the data first published in 2006\,\cite{Baker2006} (then $|\dn|\!<\!\unit[\pow{2.9}{-26}]{\ecm}$) and confirms the original result. Different EDM limits are summarized in Tab.\,\ref{tab:EDMs}, while the interested reader is referred to Ref.\,\cite{Pospelov2005,Engel2013PPNP} or a detailed overview of the interplay between different EDMs and theory\,\cite{Chupp2015PR}. 

\begin{table}%
\centering
\begin{tabular}{rclr}
\hline
EDM & limit & C.L.\ & Ref.\\ 
\hline
ThO & $d_e<\pow{8.7}{-29}\ecm$ &90\% & ~\cite{Baron2014Science}\\
\magHg & $d_{\rm Hg}<\pow{7.4}{-30}\ecm$ &95\% &~\cite{Graner2016PRL}\\
neutron & $\dn <\pow{3.0}{-26}\ecm$ &90\% & ~\cite{Pendlebury2015PRD}\\
\hline
\end{tabular}
\caption{Most relevant experimental limits on electric dipole moments.}
\label{tab:EDMs}
\end{table}

The first searches for a neutron EDM, starting in the 1950s\,\cite{Purcell1950PR,Smith1957PR}, were undertaken using thermal, and later cold, neutron beams from reactors. The last beam experiment\,\cite{Dress1977} published a limit of $\dn\!<\!\unit[\pow{3}{-24}]{\ecm}$ (C.L.90\%) in 1977 and was limited by the velocity-dependent $\boldsymbol{v\!\times\!E}$ systematic effect. At the beginning of that decade, the first experiments using ultracold neutrons\,\cite{Lushchikov1969JETPL,Shapiro1970SPU} were proposed with first results published at the end of the 1970s\,\cite{Altarev1980NuPhA}.
Today several competing collaborations around the world are pursuing new experiments to improve the sensitivity to the nEDM and by that establish a finite value or push the limit on the nEDM by up to two orders of magnitude in the next decade or so. This review is based on an earlier version\,\cite{Schmidt-Wellenburg2016LASNPA} which was published as part of the proceedings of the XI Latin American Symposium on Nuclear Physics and Applications in Medellin, Columbia. Section\,\ref{sec:ExpTech} presents the state-of-the-art methods and techniques which are used in nEDM searches with UCN, while section\,\ref{sec:Systematics} discusses the most important systematic effects. Finally, in section\,\ref{sec:WWnEDM} an overview of the worldwide activities is given.

\section{The nEDM in the Standard Model and beyond }
\subsection{Symmetries and EDMs}
\label{sec:Symmetries}
Symmetries are universally present throughout human culture and can be found throughout many concepts of art and architecture. The eightfold path of Buddhism is visualized in a wheel with an eightfold discrete symmetry. Likewise the Hindu swastika has a fourfold discrete symmetry, and thousands of visitors marvel each day at the symmetric architecture of the Taj Mahal mirrored in the reflecting pool.
At the same time it seems that the breaking of such symmetries, as for instance in the rules of the golden ratio, make a work of art or architecture even more exciting and interesting.


Symmetries and their violation also play an important role in fundamental physics. In physics the concept of symmetry is strictly defined by a transformation of an object which does not alter the object; one says that the object is invariant under this transformation. The more such transformations are allowed, the higher the degree of symmetry. As an example take again the eightfold wheel visualizing the principle virtues of Buddhism. One may mirror it along any of its spokes passing through the origin, or one may rotate it by any multiple of $\pi/4$, and it will not change its appearance as it has an eightfold rotational symmetry. A perfect circle has an even higher symmetry, as one can rotate it by any angle without changing its appearance.
As early as 1918 Emmy Noether\,\cite{Noether1918} showed that the conversation of energy, momentum and angular momentum can be directly derived from the symmetry of the laws of physic under translation in time, and under translation and rotation in space. In this context a physical law is symmetric under transformation if any transformation of the system under this operation does not alter the physical law. 
In modern particle physics three discrete symmetries, charge conjugation ({\it C}), parity inversion ({\it P}), and time reversal ({\it T}) , play an outstanding role in our understanding of nature. Their combined conservation ({\it CPT} symmetry) together with Lorentz invariance are cornerstones of today's conception of fundamental physics. 
All three of the individual symmetries are violated in the weak sector of the standard model.

\subsubsection*{Charge}
\label{sec:Charge}
The notion of charge in modern particle physics extends beyond the classical notion of electrical charge to all charge quantum number of the particle. In this sense charge conjugation is not only the reversal of the electrical charge of the object, but also of its color charge, its lepton or baryon number, its flavor quantum number and so on. Therefore, charge conjugation transforms a particle into its anti-particle by reversing all charge quantum numbers of the particle, for example the electron into the positron or the neutron ($\mrm{udd}$) into an anti-neutron ($\mrm{\bar{u}\bar{d}\bar{d}}$).\\
A system or physical process is considered as invariant under charge conjugation if the observed process is indistinguishable whether anti-particles or particles with the corresponding conjugated fields participate. In this sense the movement of an electron in an electric field is indistinguishable from a positron in an electric field of same magnitude but opposite direction.

\subsubsection*{Parity}
\begin{figure}%

	\includegraphics[width=0.90\columnwidth]{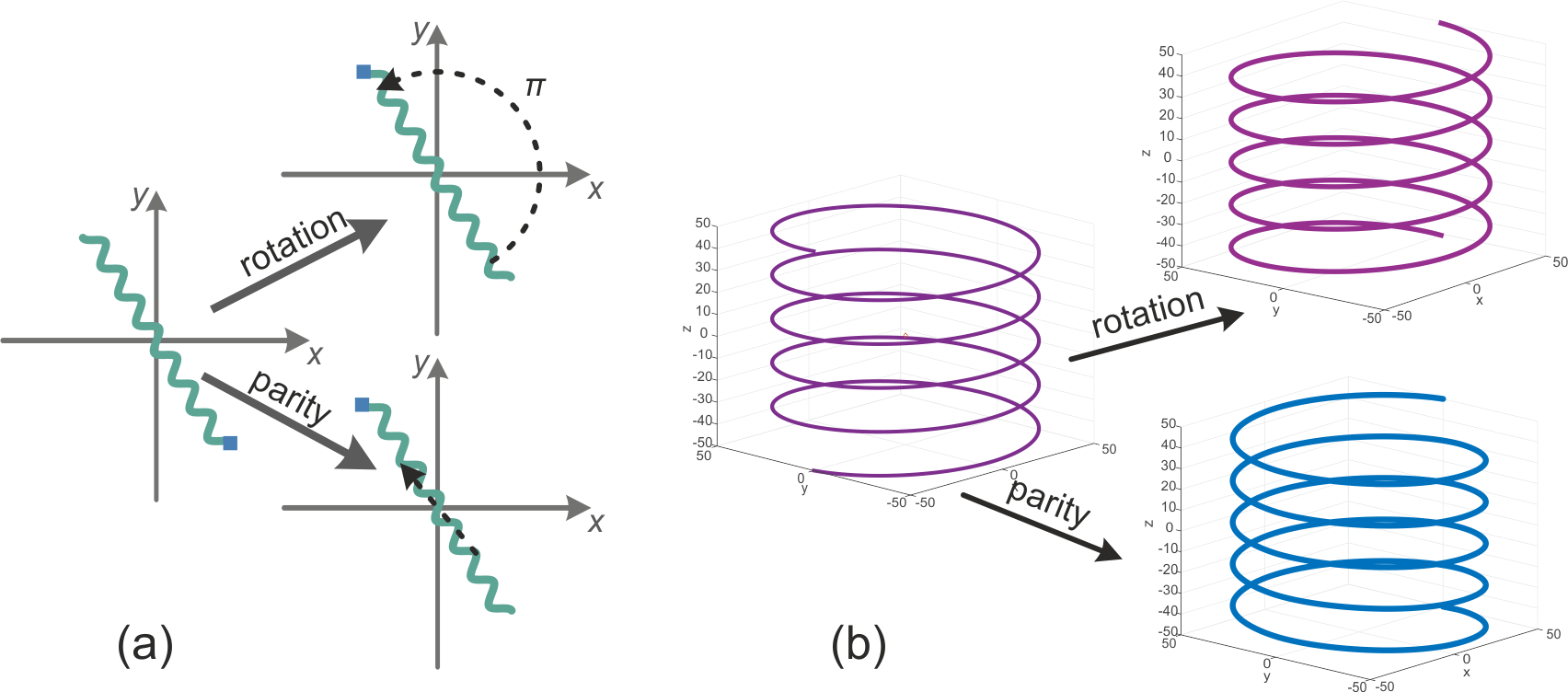}%

\caption{Illustration of parity transformation in comparison to rotational symmetries in two and three dimensions. In two and every higher even dimension the discrete parity transformation can be replaced by a set of discrete rotational transformation. (a) The discrete parity transformation ($\vec{r} \rightarrow -\vec{r}$) of all coordinates in 2 dimensions of the object is equivalent to a rotation by $\pi$ around the origin. Further the object is invariant under parity transformation, as it is indistinguishable from the original object. Note, the little blue square marks one end of the object in order to show that the object is transformed, but is not to be regarded as part of the object itself. (b) The parity transformation in 3 dimension of a screw thread ($\vec{r} \rightarrow -\vec{r}$) cannot be replaced by a series of discrete rotational transformations. Further, the sense of rotation of the screw changes. This is called the handedness. Parity transforms a right handed object into a left handed object.}
	
	\label{fig:IllustrParity}%
\end{figure}

Parity transformation is defined as the sign inversion of all space coordinates in a Cartesian coordinate system. As demonstrated in Fig.\,\ref{fig:IllustrParity}a) this leads to nothing new in two dimension where it can be replaced by a rotation around the origin by $\pi$. Whereas in three dimensions, illustrated in Fig.\,\ref{fig:IllustrParity}b), it becomes obvious that a parity transformation cannot be replaced by any number of discrete rotation. This additional symmetry can also be called the symmetry between right-handiness and left-handedness. There seems to be no obvious reason why in empty space nature would have a preference for right or left-handedness. 

\label{sec:Parity}

\subsubsection*{Time}

Time reversal symmetry indicates that a physical process which is described by a Lagrangian $\mathcal{L}(t)$ or Hamiltonian $\mathcal{H}(t)$ transforms into a process $\mathcal{L}(-t)$,$\mathcal{H}(-t)$ which again is an allowed process by physical laws. This does not mean that the actual ``arrow'' of time has to be inverted, but instead that in a video of the process it becomes indistinguishable whether the video is played forward or backward. The classical example for such processes is the elastic scattering of pool balls on a pool table. In the case of a head-on collision with total momentum transfer the moving ball will come to rest while the ball which was initially at rest will start moving with the same momentum. In this simple case it is obvious that from watching the video is is unclear whether it is played forward or backward. Still our conception of time is mostly define from seemingly irreversible processes: e.g.\ a glass falling off the table and breaking in thousand of pieces or the initial break in snooker just to mention two. In these cases it seems to be easy to distinguish whether the video is played forward or backwards. But amusingly the processes are still invariant under time reversal symmetry. Our everyday experience just tells us that it is extremely unlikely to match the initial conditions so perfectly that an observation of the seemingly backward process is possible in nature.

\label{sec:Time}


\subsubsection*{CP-violation and the nEDM}

In the electroweak sector of the SM {\it C} and {\it P} are maximally violated, recall the (V-A)-structure of the interaction, while the combined symmetry of charge and parity ({\it CP}) is conserved in most cases.
Though high-precision experiments showed that {\it CP} is also violated in rare decays of {\it K}\,\cite{Christenson1964} and \textit{B} mesons\,\cite{Abe2001PRL,Aubert2001}, an essential ingredient for baryogenesis\,\cite{Sakharov1991}. Unfortunately this known source of {\it CP}-violation is too small to explain the observed baryon asymmetry of the universe\,\cite{Riotto1999ARNPS,Morrissey2012NJP} in combination with the other known ingredients from the standard model. 

In the classical concept of an EDM described by Eq.\,\eqref{eq:dnClassic} the EDM is a vector changing its direction with parity: $P(\mbf{d}^{\rm cl}) = -\mbf{d}^{\rm cl}$. In a particle with angular momentum $\mbf{j}$ or spin $\sigma = \hbar/2$ all vectors and pseudo-vectors are either aligned or anti-aligned with the pseudo-vector of angular momentum of the particle, as the transverse component is averaged out due to rotation. Therefore one may write

\begin{equation}
	\mbf{d}^{\rm cl} = \delta_n \mbf{j} \quad \overset{P}{\longrightarrow}\quad -\mbf{d}^{\rm cl} = \delta_n \mbf{j} .
\label{eq:clHamiltonianP}
\end{equation}

\noindent Both equations can only be fulfilled if either $\mbf{d}^{\rm cl} = 0$, or parity is violated, or the angular momentum is not defining a non-degenerated state. 
However, all experiments involving neutrons are obeying to Pauli exclusion principle with a twofold degeneracy in the absence of a magnetic field. This indicates that the spin quantum number is sufficient to describe the neutron ground state. Similarly one can write for a {\it T}-transformation

\begin{equation}
	\mbf{d}^{\rm cl} = \delta_n \mbf{j} \quad \overset{T}{\longrightarrow}\quad \mbf{d}^{\rm cl} = -\delta_n \mbf{j} .
\label{eq:clHamiltonianT}
\end{equation}

\begin{figure}%
	\centering
	\includegraphics[width=0.3\columnwidth]{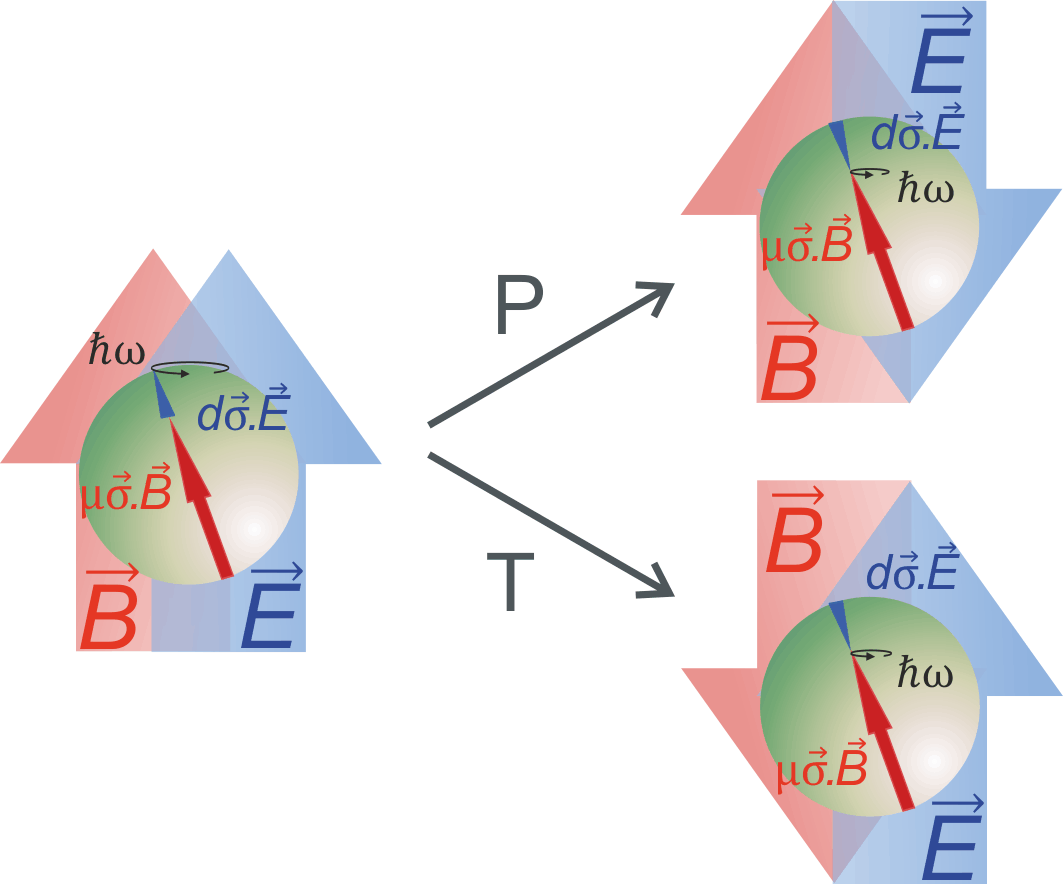}%
	\caption{Pictogram  illustrating the  {\it P} and {\it T}-violation of a nEDM in the presence of an electric and magnetic field.}%
	\label{fig:CPViolCartoon}%
\end{figure}

Which again can only be fulfilled by violating {\it T} symmetry. 
In the same way one can show in quantum mechanics that if a neutron were to have an ``electric dipole moment $\dn{}$ then, as
any vector operator in quantum mechanics, it would be connected to
the spin operator as $ \dn= \delta_\text{n} \mathbf{j}/j\hbar$, or, for $j= 1/2$ as $\dn= \delta_\text{n}\boldsymbol{\sigma}$,
where \dn{} gives the size of the EDM''\cite{Dubbers2011}\@. In the non-relativistic limit, the interaction Hamiltonian can be written as:

\begin{equation}
	H = -\frac{\hbar}{2}(\delta_\text{n}\boldsymbol{\sigma\!\cdot\!E} + \gamma_\text{n} \boldsymbol{\sigma\!\cdot\!B}),
	\label{eq:hamiltonian}
\end{equation}

\noindent where $\delta_\text{n}$ and $\gamma_\text{n}$ can be interpreted as scalar coupling strengths of the neutron spin to the electric and magnetic field. The relative sign of the two dipole coupling strengths is not defined as no electric dipole moment has yet been discovered. The magnetic coupling strength is nothing else than the gyromagnetic ratio of the neutron $\gamma_\text{n}/(2\pi)=\SI{-29.1646943(69)}{MHz\per\tesla}$\,\cite{Greene1978}, which is the ratio of the magnetic moment of the neutron $\boldsymbol{\mu_\text{n}}$ to its angular momentum $\sigma = \hbar/2$. Similarly one can introduce a gyro-electric ratio in combination with the electric dipole moment. Equation\,(\ref{eq:hamiltonian}) and Fig.\,\ref{fig:CPViolCartoon} demonstrate that the eigenstates of the Hamiltonian change when applying either a {\it T} or {\it P}- transformation to the Hamiltonian, indicating again the violation of {\it T} and {\it P}-symmetry. 
The CPT-theorem (see standard text books on field theory, e.g.\,\cite{Maggiore2005Book}) is fundamental to any modern quantum field theory and states, that any locally Lorentz-covariant field theory of a point like particle is CPT invariant. As a consequence  the observation of a \dn{} would not only indicate the breaking of time reversal symmetry but also CPV and would therefore constrain any model of CPV that attempts to explain the BAU\@. 
In any case it would manifest a new, flavor maintaining source of CPV. 

\subsection{The standard model prediction}
In the SM two sources of CPV exist: {\it i})~In the weak interaction the weak mass eigenstates of the quarks are not identical to the flavor eigenstates. Both eigenstates are interconnected via the Kobayashi-Maskawa matrix $V_\text{KM}$ which has one single phase $\delta$ which induces the observed CPV in the K and B meson decays.
	{\it ii})~The second source is the QCD vacuum polarization term, the only CP-odd term of dimension four in the SM QCD Lagrangian. For a detailed discussion see also the reviews by Pospelov and Ritz\,\cite{Pospelov2005} and by Seng\,\cite{Seng2015PRC}.

\begin{figure}%
\centering
\includegraphics[width = 0.25\columnwidth]{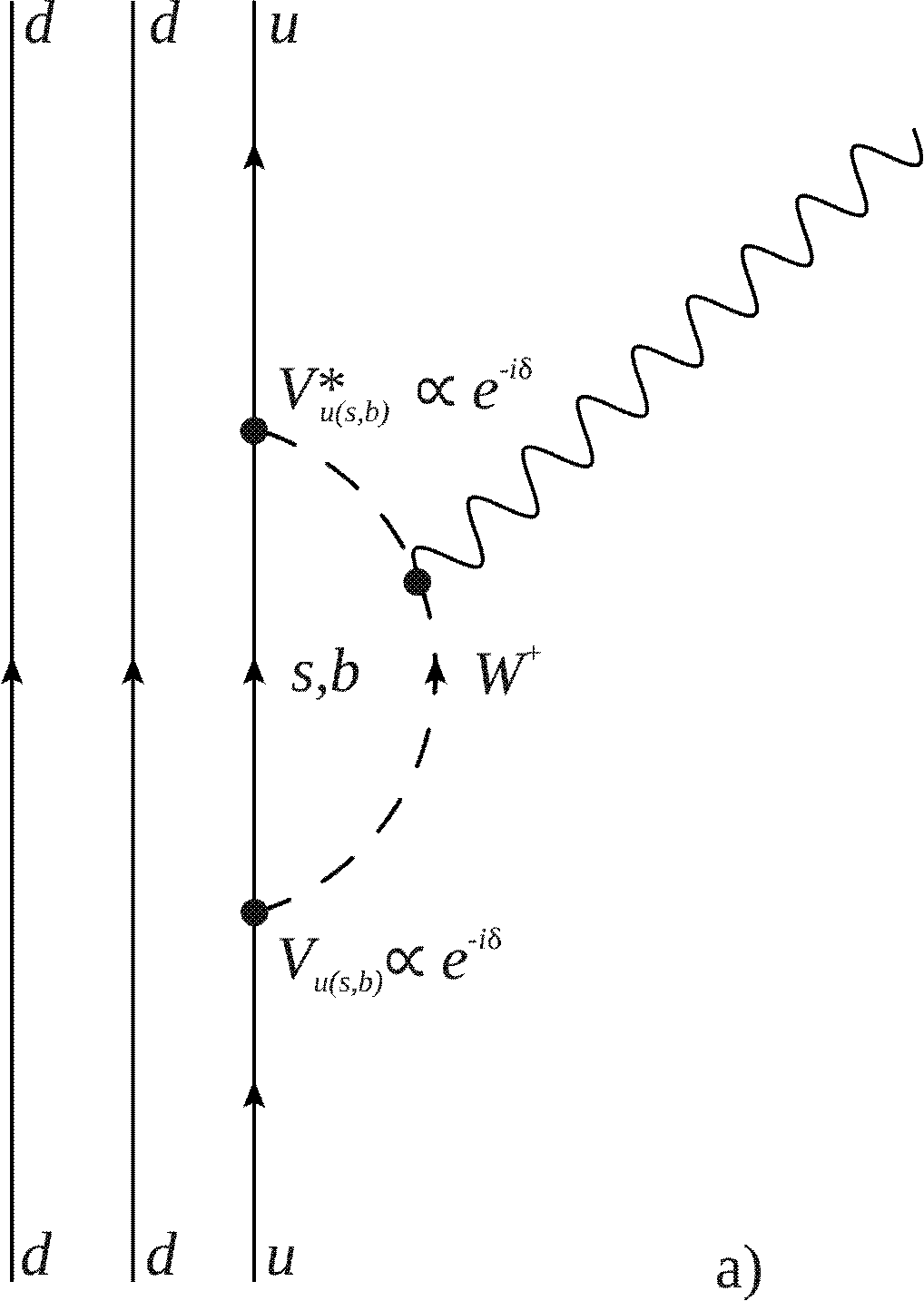}%
\hspace{3cm}
\includegraphics[width = 0.4\columnwidth]{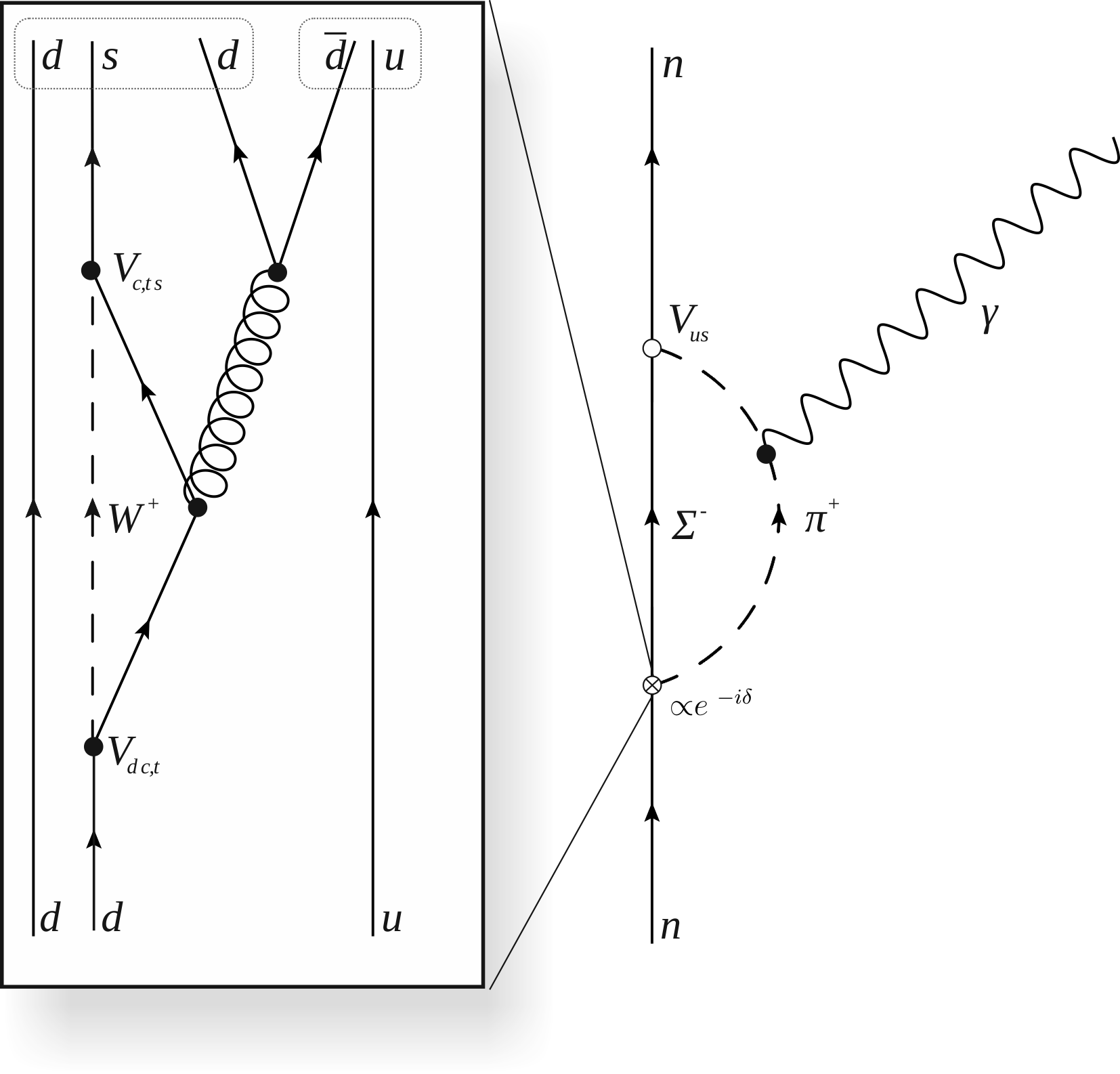}
\caption{Diagrams which involve the CPV phase of the $V_{\rm KM}$-matrix. a)~First loop contribution does not give  rise to an EDM, as the phase from the first vertex cancels in the complex conjugate second vertex. b)~This is the largest SM contribution generated by the CPV phase of the $V_{\rm KM}$-matrix. The crossed vertex, shown as insert, is a four-quark vertex known as strong penguin diagram in which not all phase contributions cancel.}%
\label{fig:WeakDiagramloops}%
\end{figure}

The Kobayashi-Maskawa matrix $V_\text{KM}$ can be written as
\begin{equation}
	V_\text{KM} = \left(\begin{array}{ccc} 
											c_{12}c_{13} & s_{12}c_{13} & s_{12}e^{-i\delta} \\
											-s_{12}c_{23}-c_{12}s_{23}s_{13}e^{-i\delta} & c_{12}c_{23}-s_{12}s_{23}s_{13}e^{-i\delta} & s_{23}\\
								s_{12}s_{23} - c_{13}c_{23}c_{13}e^{-i\delta}			&  - c_{12}s_{23}s_{13}e^{-i\delta}& c_{23}c_{13}\\
								\end{array}\right)
\label{eq:VKM}
\end{equation}
\noindent where $c_{ij} = \cos\theta_{ij}$, $s_{ij} = \sin\theta_{ij}$ and $\delta \approx \unit[1.20]{rad}$ is the CPV phase. It is impossible to write down a tree level diagram generating an electric dipole interaction of one quark of the neutron with the electric field. At the one-loop level, shown in Fig.\,\ref{fig:WeakDiagramloops}a, any phase term of a $V_{ij}$ element at one vertex will be canceled by the complex conjugated phase term at the second vertex $V_{ij}^{\ast}$. Further, Shabalin\,\cite{Shabalin1983} showed that the contributions of all second order processes to an nEDM cancel. The largest SM contribution is at the three-loop level via a strong penguin diagram\,\cite{Khriplovich1982PL} (see Fig.\,\ref{fig:WeakDiagramloops}b) which amounts to an approximate $\dn{}^\text{KM}$ of \unit[\pow{1}{-32}]{\ecm}\,\cite{Pospelov2005,Seng2015PRC}, well below current and most probable all future experimental sensitivities.

An additional source of CPV in the SM is the vacuum term of the QCD-Lagrangian:

\begin{equation}
	\mathcal{L}_\text{QCD}^\text{CPV}=\frac{g_s^2}{32\pi^2}\overline{\theta}G^a_{\mu\nu}\widetilde{G}^{\mu\nu,a},
\label{eq:QCDDim4}
\end{equation}

\noindent the only CPV dim-4 operator, where $g_s$ is the coupling constant of the strong interaction, $ \overline{\theta}$ is a phase which also includes the CPV phase of the weak interaction and $G^a_{\mu\nu}$ is the gluon field tensor. The structure of the gluon field tensor times its dual corresponds in electro-magnetism to a scalar product of $\mbf{E}\cdot\mbf{B}$ which is odd under {\it P} and {\it T} reversal. From a dimensional analysis\,\cite{Khriplovich1997Book} one can estimate the size of an nEDM generated by this term:

\begin{equation}
	\dn^\text{QCD} \approx \overline{\theta}\cdot\unit[\pow{1}{-16}]{\ecm}.
\label{eq:QCDThetaApprox}
\end{equation}

\noindent Hence, the current experimental limit on an nEDM is also a limit on $\overline{\theta} \lesssim 10^{-10}$. This is astonishing, as $\overline{\theta}$ is a phase which in principle could acquire any value between $0$ and $2\pi$. It is considered as unnatural that the value is so tiny. Possible solutions to this ``strong {\it CP} problem'' include having at least one mass-less quark, or a mechanism proposed by Peccei and Quinn\,\cite{Peccei1977,Wilczek1978,Weinberg1978}, which gives rise to the axion, a Nambu-Goldstone boson. For searches for the axion and axion like particles see, e.g., Ref.\,\cite{Baer2015}. However, if an nEDM were to be found further measurements of EDMs (proton, electron...) would be necessary to distinguish the CPV source(s) and to explain the role of the tiny $\overline{\theta}$-term.

\subsection{Generic sensitivity of an nEDM to physics beyond the SM}
\begin{figure}%
\includegraphics[width=0.6\columnwidth]{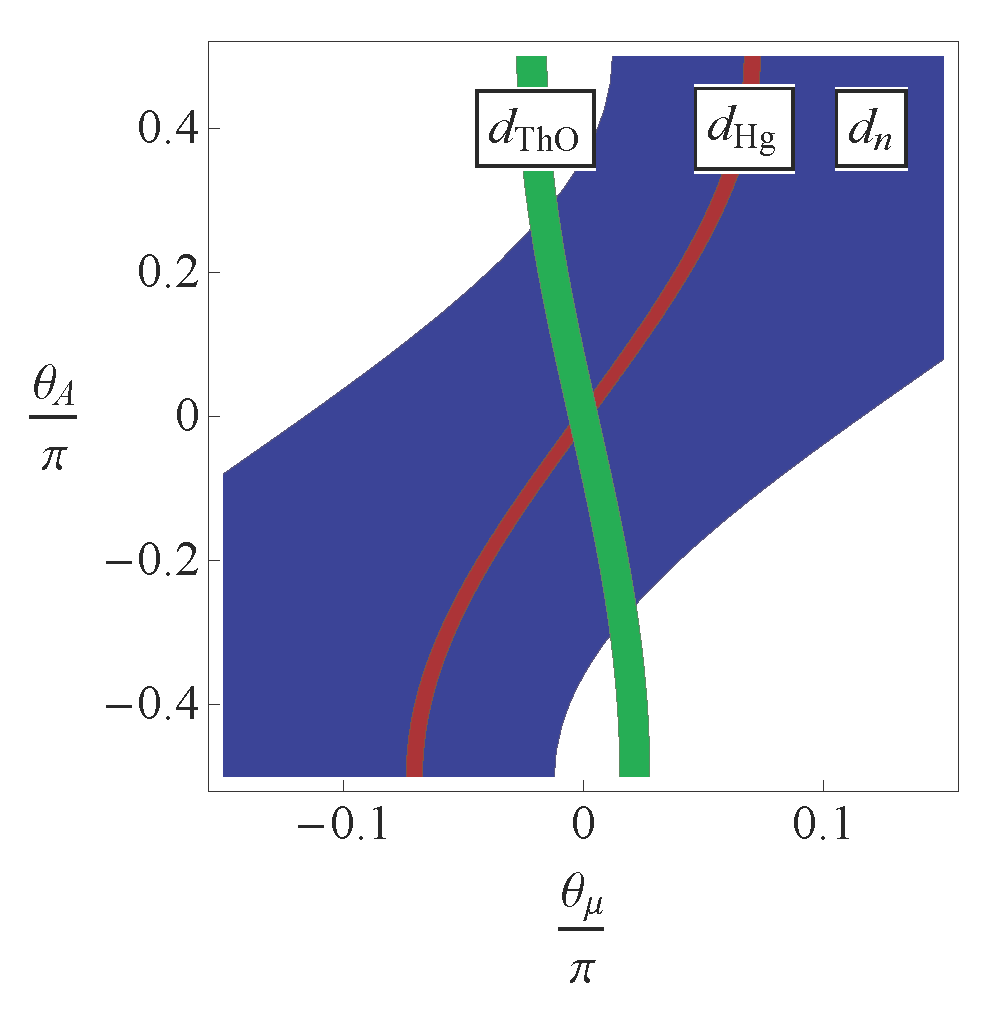}%
\caption{Constraints on two CP violating phases of a generic SUSY-model using limits from the neutron\,\cite{Pendlebury2015PRD}, Hg\,\cite{Graner2016PRL} and ThO\,\cite{Baron2014Science} (electron). The allowed phase space is already stringently reduced for a SUSY scale at \SI{2}{TeV}. Updated figure courtesy of A.~Ritz, original figure appeared in Ref.\,\cite{Pospelov2005}.}
\label{fig:SUSYCPproblem}
\end{figure}

A comprehensive summery of different BSM scenario which provide viable sources of CPV may be found in the reviews Refs\,\cite{Jaeckel2012JHEP,Engel2013PPNP}. Neglecting for a moment the contribution to an nEDM from the unnatural small $\overline{\theta}$ parameter, it is clear that any observed $\dn>\unit[10^{-30}]{\ecm}$ can only be explained by new physics. Generically most BSMs provide several CPV phases and new particles which could already contribute at the one-loop quark level to an observable nEDM\@.  A typical order of magnitude analysis from super-symmetric (SUSY) models (e.g.\,\cite{Abel2006JHEP}) gives:

\begin{equation}
		\dn \sim \left(\frac{\unit[300]{GeV}}{\Lambda_{\rm SUSY}}\right)^2\sin\phi_{\rm CP}\times10^{-24}\ecm,
\label{eq:BSMapprox}
\end{equation}

\noindent where $\phi_{\rm CP}$ represents the relevant possible CPV phases of the model and $\Lambda_{\rm SUSY}$ is the SUSY mass scale. The current experimental limit already implies that models either have to be considerably fine tuned to have a small $\phi_{\rm CP}$ or to suppress 1-loop contributions, or that the SUSY-scale is considerably above the weak scale in the range of some TeV\@. Figure\,\ref{fig:SUSYCPproblem} illustrates this conundrum known as SUSY CP problem. The authors of Ref.\,\cite{Engel2013PPNP}  generalize this SUSY approach and find similar model-independent constraint for a general BSM scale.

\section{Experimental Techniques}
\label{sec:ExpTech}
Experiments searching for the electric dipole moment of the neutron essentially try to measure the precession frequency of the neutron in a strong electric field $\omega_\text{E} = \delta_\text{n}\mbf{E}$. The current upper limit of $\dn \leq \unit[\pow{3}{-26}]{\ecm}$ (90\% C.L.) indicates that it will be necessary to measure a frequency below $\omega_\text{E}/(2\pi)=\SI{24}{nHz}$ for an electric field of $E=\SI{10}{kV/cm}$. Any magnetic field larger than \SI{0.83}{fT}, a field too small for even the best magnetically shielded rooms on Earth, would lead to a similar or larger Larmor precession frequency of the neutrons. Thus, it seems impossible to directly measure the effect of an nEDM exposed ``only'' to an electric field. Instead the neutron is exposed in addition to a well controlled magnetic field $\boldsymbol{B}$. By taking the difference between two Larmor frequencies measured in configurations where the electric field is parallel ($\omega^{\parallel}$) and anti-parallel ($\omega^{\nparallel}$) to the magnetic field we find that

\begin{eqnarray}
	\hbar\omega^{\parallel} &= & 2\left|\boldsymbol{\mu_\text{n}\!\cdot\!B}^{\parallel}+\boldsymbol{\dn\!\cdot\!E}^{\parallel}\right| \nonumber\\
	\hbar\omega^{\nparallel} &= & 2\left|\boldsymbol{\mu_\text{n}\!\cdot\!B}^{\nparallel} -\boldsymbol{\dn\!\cdot\!E}^{\nparallel} \right| \nonumber\\
	\dn &= & \frac{\hbar\left(\omega^{\parallel}-\omega^{\nparallel} \right)-2\mu_\text{n}\left(B^{\parallel}-B^{\nparallel}\right)}{2\left(E^{\parallel}-E^{\nparallel}\right)}.
\label{eq:DiffConfig}
\end{eqnarray}

\noindent In general these two measurements are either made in two adjacent volumes with opposite electric fields ($E^{\parallel}=-E^{\nparallel}$) inside the same magnetic field ($B^{\parallel}-B^{\nparallel} = 0$), or by measuring first one configuration, then changing the polarity of the electric field from $E^{\parallel}$ to $E^{\nparallel}=- E^{\parallel}$ and measuring again. In the first case it is of paramount importance to make sure that the two spatially separated measurements have the same magnetic field configuration (no or small magnetic-field gradients), while in the second case it is essential to make sure that the magnetic field is stable and/or precisely monitored in time. Both approaches are currently used or in discussion for nEDM searches.

\subsection{Ramsey's technique of separated oscillating fields}
\begin{figure}%
\centering
\subfloat{
\includegraphics[width=0.41\columnwidth]{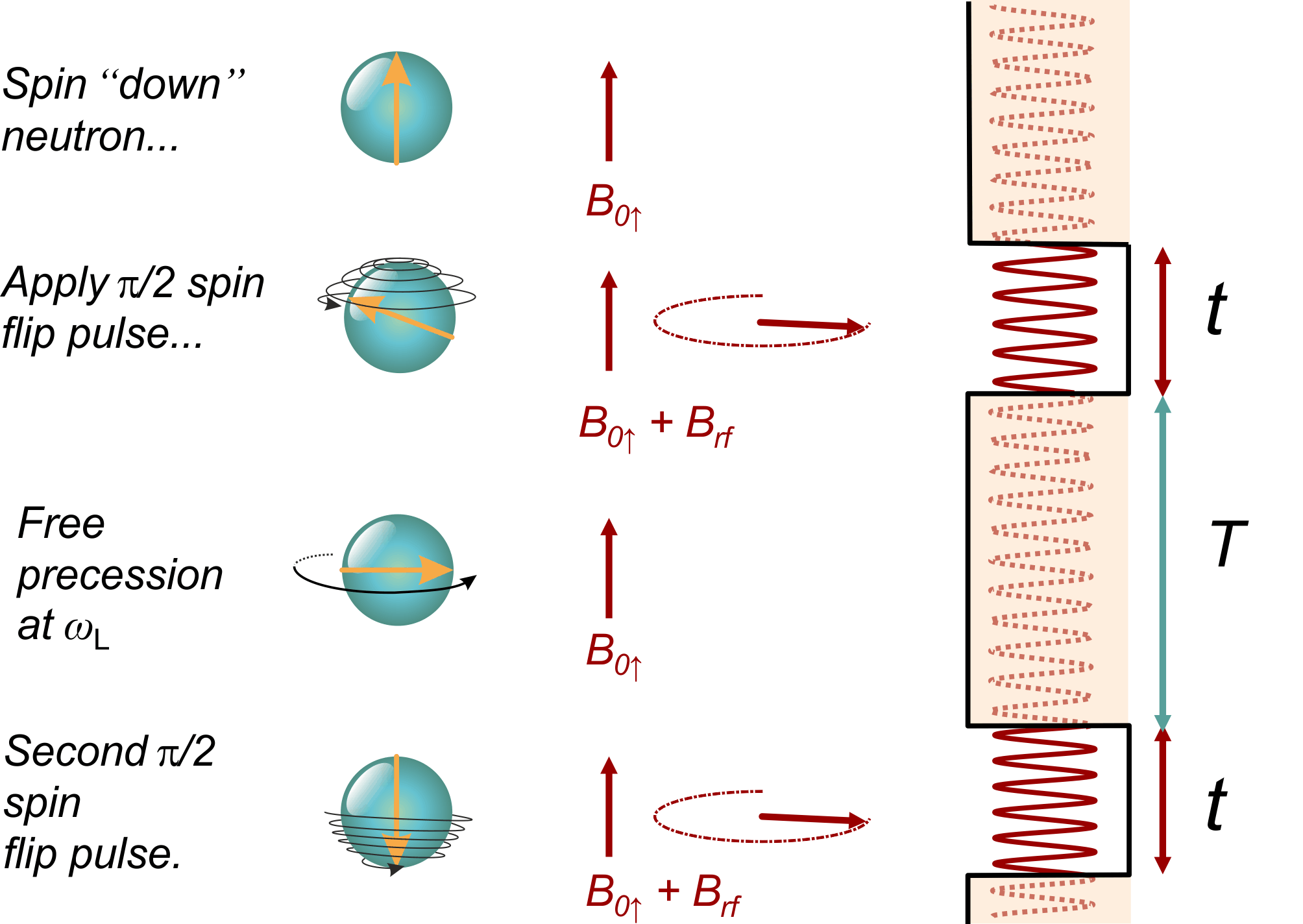}}
\subfloat{\includegraphics[width=0.40\columnwidth]{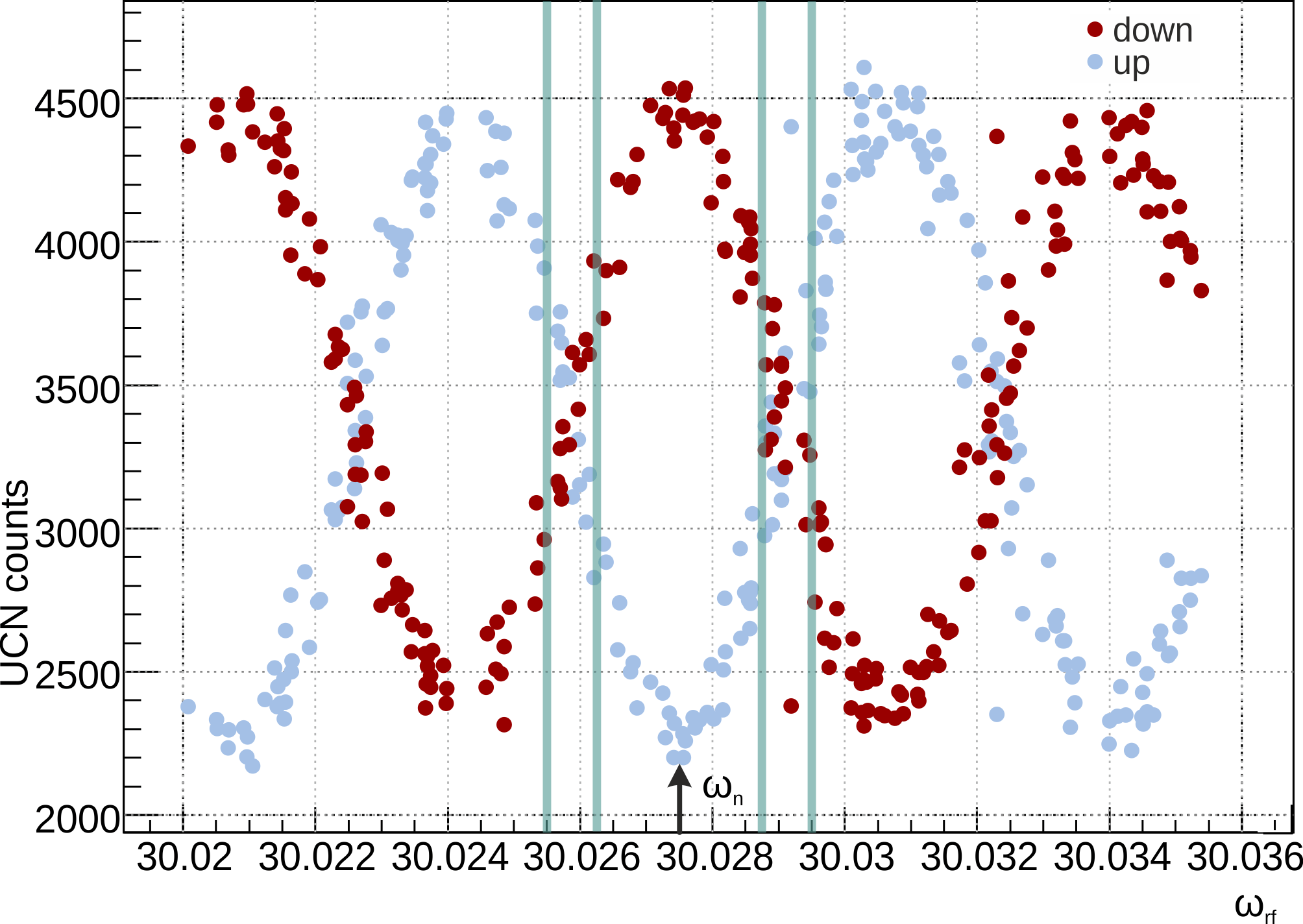}}
\caption{Ramsey's technique of separated oscillating fields. Note, that neutrons with spin down are high-field seeking neutrons indicated by the yellow arrow as the gyro magnetic ratio is negative. The scheme of the method (left), based on Fig.\,4 of Ref.\,\cite{Harris2007arXiv}, and the data plot (right) are explained in detail in the text. Blue points are UCN counted with spin up $N^{\uparrow}$, while red points are with spin down $N^{\downarrow}$ (data from the PSI-nEDM collaboration\,\cite{Baker2011}). The width at half height $\Delta \nu$ of the central fringe is approximately $1/\left(2T\right)$, and the four vertical lines indicate the working points. }%
\label{fig:RamseyMethod}%
\end{figure}

More than half a century ago Ramsey improved Rabi's resonant frequency technique to measure energy eigenstates of quantum mechanical systems by introducing a free-precession period between two spin-flipping pulses\,\cite{Ramsey1950PR}.
Figure\,\ref{fig:RamseyMethod}a) illustrates this technique, while a typical resonance scan is shown in Fig.\,\ref{fig:RamseyMethod}b). The initial state is a fully polarized, i.e.\ \stateup, ensemble of neutrons exposed to a magnetic field $B_0$. A first rotational oscillating magnetic-field pulse $B_{\rm rf} = B_1\cos\left(\omega_\text{rf}t\right)$, perpendicular to $B_0$, tips the spins into the plane orthogonal to the main magnetic field. The neutron spins then precess freely with their Larmor frequency $\omega_0$ for a duration $T$, accumulating a phase $\phi = \gamma_n B T$, before a second pulse $B_1\cos\left(\omega_\text{rf}t\right)$ in phase with the first is again applied to the neutron ensemble. The essential idea is to compare the phase $\phi$ with $\omega_{\rm rf}T$. If they are identical then $B=\omega_{\rm rf}/\gamma_n$. 

The probability to detect a neutron with a final spin state identical to its initial spin state, i.e.\ \statedo, is (see equation~(A.11) in Ref.\,\cite{Piegsa2009NIMA}):

\begin{align}
	\mathcal{P}(T,\omega_\text{rf})&=\left|\Bra{\downarrow}U(T,\omega_\text{rf})\Ket{\downarrow}\right|^2 \nonumber \\ 
	&=1-\frac{4\omega_1^2}{\Omega^2}\sin^2\left(\frac{\Omega t_{\pi/2}}{2}\right)
	\left[\frac{\Delta}{\Omega}\sin\left(\frac{\Omega  t_{\pi/2}}{2}\right)\sin\left(\frac{T\Delta}{2}\right)-\right. 
	\left.\cos\left(\frac{\Omega  t_{\pi/2}}{2}\right)\cos\left(\frac{T\Delta}{2}\right)\right]^2,
\label{eq:RamseyResonance}
\end{align}
\noindent where $U(T,\omega_\text{rf})$ is the time-evolution operator describing the pulse sequence, $\omega_1\!=\!-\gamma_\text{n}B_1$, $\Delta \!=\!\omega_\text{rf}-\omega_0$, and  $\Omega\!=\!\sqrt{\Delta^2+\omega_1^2}$. When optimized the spin-flipping pulses have exactly enough power to tip the spins by $\pi/2$, hence, the pulse length and field power fulfill the condition $\gamma_\text{n}B_1t_{\pi/2} =\pi/2$. In this case, and in the central fringe range ($\Delta \ll \omega_1$), equation\,(\ref{eq:RamseyResonance}) simplifies to:

\begin{eqnarray}
		\mathcal{P}(T,\omega_\text{rf}) &=&1-4\sin^4\frac{\pi}{4}
	\left[\frac{\Delta}{\Omega}\sin\frac{T\Delta}{2}-\cos\frac{T\Delta}{2}\right]^2 \nonumber \\
	\mathcal{P}(T,\omega_\text{rf}) &\approx& 1- \cos^2\frac{T\Delta}{2} \nonumber \\
	\mathcal{P}(T,\omega_\text{rf}) &=& \frac{1}{2}\left(1 -\cos\left(T\Delta\right)\right).
\label{eq:CosineApproximation}
\end{eqnarray}

\noindent In a real measurement with $N$ neutrons inside a large magnetic field region this becomes

\begin{equation}
	N^{\downarrow/\uparrow} = \frac{N}{2}\left\{1 \mp \alpha(T)\cos\left[ \left(\omega_\text{rf} - \gamma_\text{n}B_0\right)\cdot\left(T+\frac{4t_{\pi/2}}{\pi}\right)\right]\right\},
\label{eq:UCNRamseyFormula}
\end{equation}
\noindent where $\alpha(T)$ is the visibility of the central fringe taking into account all depolarization effects\,\cite{Afach2015PRD} and $N^{\downarrow/\uparrow}$ are the neutrons with spin down (high-field seeking) and spin up (low-field seeking). The term  $\frac{4t_{\pi/2}}{\pi}$ is necessary to account for field inhomogeneities of $B_1$ and $B_0$ which become relevant when the pulse length $t_{\pi/2}$ is finite\,\cite{Slichter1990book}. 

A Ramsey interference pattern, as shown in Fig.\,\ref{fig:RamseyMethod}b), can be recorded by scanning $\omega_\text{rf}$ while keeping all other conditions constant. The magnetic field $B$ measured by the neutrons then fulfills the resonance condition of the central fringe with $\omega_\text{rf} = \gamma_\text{n}B$.
This procedure is further optimized in searches for an nEDM by only measuring at four points close to highest sensitivity, the working points.
The neutron Larmor frequency is then obtained by fitting expression\,(\ref{eq:UCNRamseyFormula}) to the data.
Separate fits are performed for different electric-field and magnetic-field configurations.
Taking the difference of these Larmor frequencies gives access to the electric dipole moment:

\begin{equation}
		\dn = \frac{\hbar (\omega_0^{\parallel}-\omega_0^{\nparallel})}{2(E^{\parallel} - E^{\nparallel})} = \frac{\hbar \Delta\omega}{4E},
\label{eq:nEDM-simple}
\end{equation}

\noindent using equation\,(\ref{eq:DiffConfig}) and assuming no differences in the magnetic field and $E=E^{\parallel} = -E^{\nparallel}$. The statistical sensitivity with which a frequency can be measured in a time $t$ can be deduced from equation\,(\ref{eq:UCNRamseyFormula}):

\begin{equation}
	\sigma\left(\omega_0\right) \approx \frac{1}{\alpha T \sqrt{\dot{N}t}},
\label{eq:SensitivityFrequency}
\end{equation}

\noindent where $\dot{N} = \langle N \rangle \cdot f$ is the neutron current with $ \langle N \rangle$ the average number of neutrons detected in one measurement and $f$ the repetition rate. This translates into a statistical sensitivity for the detection of an nEDM of

\begin{equation}
	\sigma\left(\dn\right) \approx \frac{\hbar\sqrt{2}}{4\alpha T E \sqrt{\dot{N}t/2}}
	= \frac{\hbar}{2\alpha T E \sqrt{\dot{N}t }},
\label{eq:SensitivityEDM}
\end{equation}

\noindent  using equation\,(\ref{eq:nEDM-simple}). The additional factor $1/\!\sqrt{2}$ in the denominator accounts for the fact that it needs two measurements to take a difference.

\subsection{Ultracold neutrons}
\subsubsection{Properties of UCN}
From a neutron current point of view it becomes immediately clear that $T\!\sqrt{\dot{N}}$ needs to be maximized for the highest sensitivity. One possible solution, proposed in 1960 by Shapiro\,\cite{Lushchikov1969JETPL,Shapiro1970SPU}, is to increase $T$ by using neutrons that can be stored within vacuum chambers made of adequate materials. Such ultracold neutrons (UCN) are reflected under any incident angle by the neutron optical potential

\begin{equation}
	V = \frac{2\pi\hbar^2}{m_\text{n}}\mathcal{N}b,
\label{eq:NeutronOptPot}
\end{equation}

\noindent where $m_\text{n}$ is the neutron mass, $\mathcal{N}$ the nucleon density and $b$ the neutron scattering length of a given material. Well suited materials for storage are, e.g., $^{58}$Ni, diamond-like carbon (DLC), NiMo(85/15), or BeO/Be which all have neutron optical potentials in the range  \SIrange{200}{350}{neV}.
These potentials also define the maximum kinetic energy of UCN\@, which is in the same range as the gravitational potential energy of neutrons: $mgh\approx h
\cdot\unit[1.025]{neV/cm}$. Furthermore, strong magnetic fields can be used to polarize or store UCN, as the magnetic potential energy $\boldsymbol{\mu}\cdot\boldsymbol{B} \approx \pm \mbf{B}\cdot\unit[60]{neV/T}$. This means, a magnetic field of \SI{5}{T} creates a potential barrier of \SI{300}{neV} for low-field seekers while high-field seekers are attracted. 

For the search of an nEDM with UCN the ultimate goal is to confine UCN as long as possible within a closed material cell. Several imperfections of the material cell will effectively reduce the number of UCN of energy $E$ detected after a time $t$ to $N_0(E)\exp\left(-t/\tau_{\rm s}(E)\right)$. The storage time constant $\tau_{\rm s}(E)$ is the inverse of the sum of all loss rates:

\begin{equation}
		\frac{1}{\tau_{\rm s}(E)} = \Gamma(E) = \Gamma_{\beta}+\Gamma_{\eta}(E)+\Gamma_{\rm gap}(E) + \Gamma_{\rm gas}(E).
\label{eq:UCN_lossrates}
\end{equation}

\noindent A free neutron is not stable and decays with a rate of $\Gamma_{\beta}\approx \SI{1/880}{s^{-1}}$. The energy dependent losses due to nuclear absorption and gaps in the surface can be estimated to be

\[
\Gamma_{\eta}(E) = \frac{\mu(E)v(E)}{\lambda(E)} \quad\quad \Gamma_{\eta}(E) = \frac{v(E)S_{\rm gap}}{\lambda(E)S},
\]

\noindent where $\mu(E)$ is the energy dependent loss per bounce probability, typically in the range of some $\pow{1}{-4}$, and $S/S_{\rm gap}$ is the ratio of cell surface to surface of gaps. The last loss rate $\Gamma_{\rm gas}(E)$ from collisions with rest gas molecules can be neglected for a typical vacuum conditions. Hence, the number of UCN $N(t)$ remaining after a time $t$ inside the volume is:

\begin{equation}
		N(t) = \frac{\int {\rm d}E~ N_0(E)\exp\left(-\Gamma(E)t\right)}{\int {\rm d}E~ N_0(E)},
\label{eq:UCN_ExpDecay}
\end{equation}

\noindent where $N_0(E)$ is the energy spectrum of UCN directly after closing the door of the storage cell. However, very often the energy spectrum is not very well known and the measured data can be described  sufficiently well using a double-exponential decay:

\begin{equation}
		N(t) = N_{\rm f}\exp(-\Gamma_{\rm f}t)+N_{\rm s}\exp(-\Gamma_{\rm s}t),
\label{eq:UCN_DoubleExpDecay}
\end{equation}

\noindent where the index `f' describes a fast and the index `s' a slow loss from the storage volume. Good quality storage volumes for searches for an nEDM report values larger than $1/\Gamma_{\rm f} > \SI{60}{s}$ and $1/\Gamma_{\rm f} > \SI{180}{s}$. For a complete review of UCN physics please refer to the books\,\cite{Golub1991,Ignatovich1990}.

\subsubsection{Production of UCN}
Free neutrons are typically produced in research reactors by nuclear fission or in spallation sources where a high-power proton beam is incident on a neutron rich, often stable element (e.g.\ mercury or lead). These neutrons are then cooled by a heavy or normal water moderator to room-temperature. The energy distribution can be very well approximated by a Maxwell-Boltzmann distribution for slightly higher energies. To further shift the energy spectrum a moderator at liquid hydrogen temperature is used as source of cold neutrons.
Any cold neutron source has already a significant number of UCN in the low-energy tail of its distribution. The extraction through aluminum windows and long guides significantly reduces the amount of available UCN outside the biological-protection.

 These problems were beautifully circumvented by the conception and design of a phase-space converter, the UCN-turbine (instrument PF2) at the Institut Laue-Langevin, which Doppler-shifts very cold neutrons to the UCN regime\,\cite{Steyerl1975NIM,Steyerl1986PLA}. Until now, this instrument remained the workhorse and benchmark in UCN-physics.
All other current and next-generation UCN sources\,\cite{Trinks2000NIMA,Saunders2013RSI,Lauss2014,Piegsa2014,Serebrov2014TPL,Masuda2015JPS,Zimmer2015PRC} are based on the superthermal concept proposed by Pendlebury and Golub in 1975\,\cite{Golub1975,Golub1977}. The principal idea is to use collective excitations of the conversion medium to down-scatter neutrons from higher energies to the UCN energy regime\,\cite{Kirch2010}. For this process two materials are of main interest: superfluid helium (He-II), using the phonon-roton excitations with a relatively feeble production rate but profiting from a zero absorption cross-section\,\cite{Korobkina2002,Schmidt-Wellenburg2009}, and solid deuterium (sD$_2$) which has a broad range of excitations leading to a high conversion rate while the finite absorption cross-section reduces the effective layer from which a UCN can escape the material\,\cite{Atchison2007PRL}. Both methods can be adapted in such a way that in principle it should be possible to build UCN sources with greatly increased UCN densities, compared to today's standards\,\cite{Kirch2014FPUA}, delivered to experiments at room temperature. An alternative approach are cryogenic nEDM experiments, that intend to avoid transport losses on the path source to room-temperature precession cell altogether, which are discussed in Sec.\,\ref{sec:CryogenicApproach}.

\subsubsection{Detection of UCN}
At the end of the Ramsey cycle the number of neutrons with spin up $N^{\uparrow}$ (low-field seeking) and spin down $N^{\downarrow}$ (high-field seeking) have to be counted, before they can be fitted to expression\,\eqref{eq:UCNRamseyFormula}. For the discrimination between high and low-field seeking neutrons one can use the high magnetic field inside a thin magnetized iron layer $B_{\rm Fe} \approx \SI{2}{T}$ on an aluminum foil. If such a foil is placed inside a sufficiently strong magnetic field from permanent magnets with a flux return yoke, the low-field seeking neutrons will see a potential step of $V_{\rm F} = V_{\rm Fe} + \SI{2}{T}\!\cdot\!\SI{60}{neV/T} = \SI{330}{neV}$, while the potential step for the high-field seeking UCN is reduced to \SI{60}{neV} ($V_\text{Fe} = \SI{210}{neV}$). 
Placed approximately \SI{1}{\meter} below the precession cell, such a foil will reflect all low-field seeking (spin up) UCN, while letting high-field seeking (spin down) pass. In combination with an efficient adiabatic spin-flipper\,\cite{Luschikov1984NIMA,Geltenbort2009NIMA} placed in front of such a foil it is possible to invert the spin-state of the neutrons incident on the foil and detect both spin states. The spectrometer from Ref.\,\cite{Baker2014,Pendlebury2015PRD} used one foil and one spin-flipper that was turned on and off during the counting period for spin-analysis. An alternative approach with two detection beam lines each equipped with one foil and one spin-flipper for each precession cell was used in Ref.\,\cite{Serebrov2015PRC} and is currently employed for the single precession cell setup at PSI\,\cite{Afach2015EPJA}. Once the neutron has passed the spin-analysis stage it is detected in a UCN detector.

The neutron, being an uncharged particle, cannot be detected with conventional particle detectors which are based on ionization processes. While high energetic neutrons can be measured by recoil protons from hydrogen-dense materials, low energetic neutrons $E<\unit[25]{meV}$ are best detected exploiting nuclear reactions . Once a neutron is captured by a suitable nuclei a conventional detector may detect the charged particle emitted by the prompt nuclear reaction. Any impulse or energy information of the incident neutron is lost during this reaction, hence a UCN detector is a simple counter. Table\,\ref{tab:DetectionIsotopes} lists the three most widely used isotopes with a large absorption cross section and suitable charged reaction partners. 

\begin{table}%
\begin{tabular}{l|crl}
	Isotope	 & $\sigma_{\rm a}$ (barn) & \multicolumn{2}{c}{Nuclear reaction}  \\ \hline \hline
		\tHe	&  5330 &  ${\rm \tHe +n }~\longrightarrow$ &${\rm ^3H+ p + \SI{764}{keV}}$ \\
		$^6$Li & 940 & $\rm ^6Li + n~\longrightarrow$ &${\rm ^3H + \alpha +  \SI{4.78}{MeV}}$ \\
		{$^{10}$B} & {3835} &$\rm ^{10}B + n~\longrightarrow$ &$ \rm {^7Li} + \alpha + \SI{2.79}{MeV}~(\SI{6}{\%})$\\
			&	 & $\longrightarrow $& ${\rm ^7Li^{\ast} + \alpha + \SI{2.31}{MeV}}~(\SI{94}{\%})$\\ 
			&	 & &~$\hookrightarrow  {\rm ^7Li + \gamma + \SI{0.48}{MeV} }$ \\
			\hline
\end{tabular}
\caption{List of typically-used isotopes in UCN detectors with absorption cross section (for $\lambda=\SI{1.8}{\angstrom}$) and nuclear reaction\,\cite{NeutronDataBooklet}.}
\label{tab:DetectionIsotopes}
\end{table}

In the past \tHe-gas detectors were the most commonly used detectors in UCN physics. As their signal length is rather long (\SI{2}{\micro\second}) these detector face serious dead-time when operated at new and future UCN sources with count rates approaching \SI{1}{\mega\hertz}. Further, the price of \tHe{} has steadily risen in the last years due to the shortage in supply.
These prospects triggered several studies of alternatives using $^6$Li and $^{10}$B as converter isotope, in the last decade. In particular designs have been investigated using gaseous detection\,\cite{Morris2009NIMA,Klein2011NIMA,Salvat2012NIMA}, silicon state detection\,\cite{Baker2003NIMA,Lasakov2005NIMA,Lauer2011EPJA}, $^6$Li-doped glass scintillators\,\cite{Ban2009NIMA,Goeltl2013EPJA,Jamieson2015NIMA,Ban2016EPJA} or a $^{10}$B/ZnS(Ag) multilayer\,\cite{Wang2015NIMA}. 

\subsection{Statistical sensitivity}
The statistical sensitivity also depends on the transverse spin coherence time $T_2^{\ast}$ of the precessing spins and the UCN losses, as described in equation\,\eqref{eq:UCN_lossrates}, inside the precession chamber. Including these into equation\,\eqref{eq:SensitivityEDM} yields for UCN

\begin{equation}
		\sigma\left(\dn\right) = \frac{\hbar}{2\alpha_0\exp\left(-\frac{T}{T_2^{\ast}}\right) T E \sqrt{\langle N(f^{-1})\rangle ft}}.
\label{eq:nEDM_SensitivityUCN}
\end{equation}

For a general analysis we simplify equation\,\eqref{eq:UCN_ExpDecay} to a single exponential decay described by $\langle N(f^{-1}) \rangle=N_0\exp(-t/\overline{\tau})$. The minimum of equation\,\eqref{eq:nEDM_SensitivityUCN} is then at 

\[
	T = \frac{2T_2^{\ast}\overline{\tau}}{T_2^{\ast}+2\overline{\tau}}.
\]
Hence the best attainable sensitivity for a given electric field $E$, initial polarization $\alpha_0$ and a neutron current of $\dot{N}=N_0f$ is

\begin{equation}
	\sigma_{\rm min}(\dn) = \frac{\hbar}{4\alpha_0E\sqrt{N_0ft}}\frac{\exp\left(\frac{2\overline{\tau}}{T_2^{\ast}+2\overline{\tau}}\right)(T_2^{\ast}+2\overline{\tau})}{T_2^{\ast}\overline{\tau}\sqrt{\exp\left(-\frac{2T_2^{\ast}}{T_2^{\ast}+2\overline{\tau}}\right)}}.
\label{eq:nEDM_BestSensitivityUCN}
\end{equation}

In the current nEDM-experiment at PSI, on average 14\,000 UCN are counted every $1/f=\SI{300}{s}$, after a free precession period of $T= \unit[180]{s}$. The measured double exponential decay constants are $t_{\rm f} \approx \SI{70}{s}$ and  $t_{\rm s}\approx \SI{300}{s}$, which yields for a single exponential approximation $\dot{N}=\SI{140}{s^{-1}}$ and $\overline{\tau}=\SI{163}{s}$. The transverse depolarization time is regularly above $T_2^{\ast}=\SI{1000}{s}$ while the initial polarization is close to 0.86. With an average electric field of $|E|=\SI{11}{kV/cm}$ this gives a sensitivity of 

\begin{equation}
	\sigma(\dn(t)) = t^{-1/2}\cdot\unit[\pow{3.25}{-23}]{\ecm/\sqrt{Hz}},
\label{eq:nEDM_BestSensitivityUCN_Value}
\end{equation}

\noindent with $t$ the effective measurement time. Table\,\ref{tab:Comparison} sets this value in relation to statistical shot-noise limits of other methods to search for an nEDM.

\begin{table}
    \begin{tabular}{rccccccc}
               & T            & N-current  & alpha & E-field  & $T_2^{\ast}$   & $\overline{\tau}$  & shot-noise limit \\
		Method 	 & (s)          & ($s^{-1}$) &       & (kV/cm)  & (s)            & (s)                & (${\rm \ecm/\sqrt{Hz}}$) \\
		\hline
    last beam & 0.015       & $\pow{1.0}{6}$    & 0.80  & 100   & ~    & ~   & \pow{8670}{-25}  \\
    new beam  & 0.1         & $\pow{4.0}{8}$    & 0.75  & 100   & ~    & ~   &   \pow{22}{-25}  \\
    RT UCN    & 246 		 & $\pow{1.4}{2}$  & 0.86  & 11    & 1000 & 160 &   \pow{325}{-25}  \\
    future RT UCN & 333		   & $\pow{2.8}{3}$  & 0.95  & 11    & 2000 & 200 & \pow{48.5}{-25}  \\
    cryo-UCN  & 570 		   & $\pow{2.3}{4}$  & 0.99  & 50    & 2000 & 440 &  \pow{1}{-25}    \\
		\hline
    \end{tabular}
		\caption{Comparison of the shot-noise limit for different methods. Experiments with ultracold neutron at room temperature (RT) will profit from better UCN sources and a better controlled magnetic environment. The shot-noise limit will further decrease when going to a cryogenic setup within superfluid helium. It has been shown that in such a case the electric field can be up to ten times higher and UCN transport losses may be reduced or even become obsolete when producing UCN and measuring the spin precession within the same volume is feasible. The future beam experiment will mainly profit from high counting statistics and a ten time higher electric field which is possible as no insulator will be needed to separate the electrodes. The most important systematic effect, $\mbf{v}\times \mbf{E}$ effect, shall be controlled by exploiting the pulse structure which permits to use a TOF-method to eliminate any $\mbf{v}$-dependency. Please refer to Tab.\,\ref{tab:WWSearches} for a comprehensive overview of current ongoing and proposed searches.}
		\label{tab:Comparison}
\end{table}

\subsection{Cryogenic approach}
\label{sec:CryogenicApproach}
In many ways a completely cryogenic experiment is considered as the ultimate drive to highest sensitivity. The essential objective is to perform the entire experiment inside He-II: UCN production and polarization, Ramsey-type measurement, spin-analysis and detection. This would allow optimal exploitation of the high UCN density of a He-II based source in combination with a 5-10 times higher electric field than in vacuum\,\cite{Ito2014arXiv}. 

Ultracold neutron sources based on He-II converters at temperatures below \SI{0.5}{K} have a saturation density of $\rho_{\infty} = \tau P$, where the storage lifetime is essentially limited by the neutron beta decay $\tau_{\beta} \approx \SI{880}{s}$ and $P=\diff\phi/\diff\lambda \cdot \SI{5e-9}{\angstrom\per\cubic\cm}$ is the production for a differential cold neutron flux at \SI{0.89}{\angstrom}\,\cite{Schmidt-Wellenburg2009}. 
Storage times of $1/\Gamma_{\rm f} = \SI{97(13)}{s}$ and  $1/\Gamma_{\rm s} = \SI{580(30)}{s}$ extracted with a two-exponential fit were reported from tests of a cyrogenic storage volume of \SI{3}{\liter} coated with deuterated polystyrene\,\cite{Leung2016Mainz}. Combining this with a flux from a CN beam like PF1 at ILL\,\cite{Abele2006} results in a saturation density of $\rho_{\infty}
\approx \SI{4500}{\per\cubic\cm}$.

The cryo-EDM collaboration\,\cite{Baker2010JPhCS} followed the classical UCN storage and measurement approach by connecting a cyrogenic precession cell to a He-II UCN source. The UCN were transported from the source in beryllium coated guides immersed in the \SI{0.5}{K} cold liquid helium to the precession cell were the electric and magnetic field were to be applied. After the Ramsey cycle the precession cell was connected to a cryogenic UCN detection system\,\cite{Baker2003NIMA}. The funding for the project was stopped in 2014 after persistent technical difficulties and an accumulated delay of several years.
 
An even more radical approach is followed by the SNS-EDM collaboration\,\cite{Golub1994PhR} with the intention to avoid UCN transport altogether. In this case the detection of the neutron spin precession is realized using a dilute solution of spin polarized \tHe{} in He-II. The absorption of a neutron only occurs when the total spin is zero. The reaction products, a proton and a tritium nuclei (see Tab.\,\ref{tab:DetectionIsotopes}), produce ultra violet light in He-II which can be detected by photomultiplier tubes. Therefore it becomes possible to use the intensity of scintillation light, after a classical Ramsey cycle, as measure of spin asymmetry. Alternatively it might be possible to measure the scintillation light from the beating of the two precession frequencies of \tHe{} and neutron or to employ the `dressed spin' method in the presence of a strong oscillating magnetic field\,\cite{Esler2007PRC}.

Combining these cryogenic prospects in equation\,\ref{eq:nEDM_SensitivityUCN} with $\langle N(T)\rangle f = \SI{2.3e4}{\per\second}$, $t_f=1/\Gamma=\SI{440}{s}$ and $T_2^{\ast} = \SI{2000}{s}$ yield a shot noise limit of \SI{1e-25}{\elementarycharge \cm \persqrthz}, which seems by today's technical standard as the ultimate feasible sensitivity. After two years of operation this experiment could reach a sensitivity below \SI{2e-29}{\elementarycharge \cm}.


%
 %

\section{Systematic effects}
\label{sec:Systematics}
A substantial increase in counting statistics will immediately require the control of all systematic effects of a similar level. Two different types of systematic effects may be identified: i)~effects correlated to the change of the field configuration, for example a leakage current, mimicking the signal of an nEDM; and ii)~effects of a stochastic nature such as random magnetic field drifts, which reduce the attainable sensitivity. This is visible from the last line of equation\,(\ref{eq:DiffConfig}):

\begin{equation}
	\dn =  \frac{\hbar\left(\omega^{\parallel}-\omega^{\nparallel} \right)-2\mu_\text{n}\left(B^{\parallel}-B^{\nparallel}\right)}{2\left(E^{\parallel}-E^{\nparallel}\right)},
\label{eq:DiffConfig2}
\end{equation}

\noindent with $B^{\parallel} = B_0+\delta B^{\parallel} + \delta B(E^{\parallel})$ not necessarily  equal to $B^{\nparallel} = B_0+\delta B^{\nparallel} + \delta B(E^{\nparallel})$. Here $\delta B$ describes any uncorrelated drifts of the magnetic field and $\delta B(E)$ describes magnetic fields caused by the electric field.

\subsection{Requirements for the stability of the magnetic field}

\begin{figure}%
	\includegraphics[width=0.9\columnwidth]{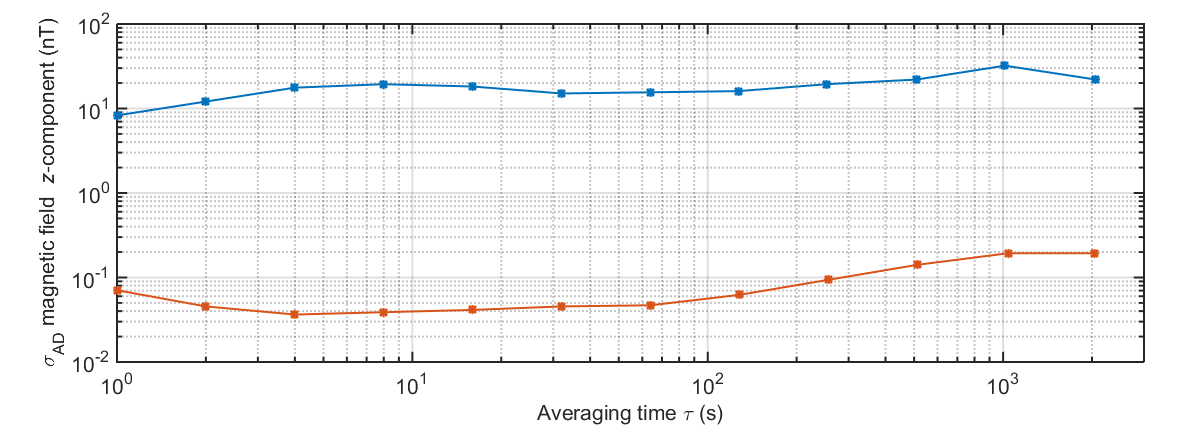}%
		\caption{Allan deviation of the $z$-component of the environmental magnetic field at the location of PSI's nEDM apparatus. The field is measured using 10 fluxgate field sensors placed in the near vicinity of the four-layer magnetic shield. The difference between night time (solid lines) and day time is significant. Figure taken from reference\,\cite{Afach2014JAP}.}%
\label{fig:AD}%
\end{figure}

In order not to degrade an initial statistical sensitivity $\sigma(\dn)$ by random magnetic field drifts it is necessary to guarantee that

\begin{equation}
	\delta B = \left(\delta B^{\parallel}-\delta B^{\nparallel}\right) \ll \frac{2\left|E\right|\sigma(\dn)}{\mu_\text{n}},
\label{eq:MagneticFieldStability}
\end{equation}

\noindent for two measurements with inverted electric field configuration. This results in a required field stability of much better than \SI{25}{fT} over a period of \SI{9}{\hour}, for a field reversal every \SI{4.5}{h} and a neutron statistical sensitivity of $\sigma(\dn)=\pow{1}{-24}\ecm$ per Ramsey cycle. The stability of the magnetic field may be best characterized by an Allan deviation (AD) of the magnetic field measured using an auxiliary magnetometer. The AD is defined as a function of averaging time $\tau$ by:

\begin{equation}
	\sigma_{\rm AD}\left(\tau_{\rm AD})\right) = \sqrt{\frac{1}{2\left(N_{\rm AD}-1\right)}\sum^{N_{\rm AD}-1}_{i=1}\left(B_i(\tau_{\rm AD})-B_{i+1}(\tau_{\rm AD})\right)^2},
\label{eq:AD}
\end{equation}

\noindent where $N_{\rm AD}=T_{\rm AD}/\tau_{\rm AD}$ is the number of non-overlapping samples of magnetic field values $B_i(\tau_{\rm AD})$ each averaged over a time $\tau_{\rm AD}$ in a measurement of total duration $T_{\rm AD}$. Figure\,\ref{fig:AD} shows an AD for several averaging times $\tau_{\rm AD}$ taken from reference\,\cite{Afach2014JAP} describing a system to stabilize the magnetic field for the nEDM search at PSI\@. The best field stability is reached during night times for averaging times of not more than \SI{100}{s}, but even then the typical variation is of the order \SI{100}{pT}, far too large for any competitive nEDM search. For this reason all current and proposed future experiments use a multilayer magnetic shield made of a high-permeability ($\mu$)
NiFe alloy which may have a static shielding factor of better than $100\,000$\,\cite{Sumner1987JPhD,Altarev2014RSI}.
In addition, an active surrounding field compensation system\,\cite{Afach2014JAP} based on several fluxgate magnetometer in close vicinity of the shield may be used to keep the shield in a stable magnetic field. Such a scheme uses a feedback algorithm to actively control a set of large coils around the magnetic shield to correct for magnetic-field changes.

\begin{figure}%
	\includegraphics[width=0.9\columnwidth]{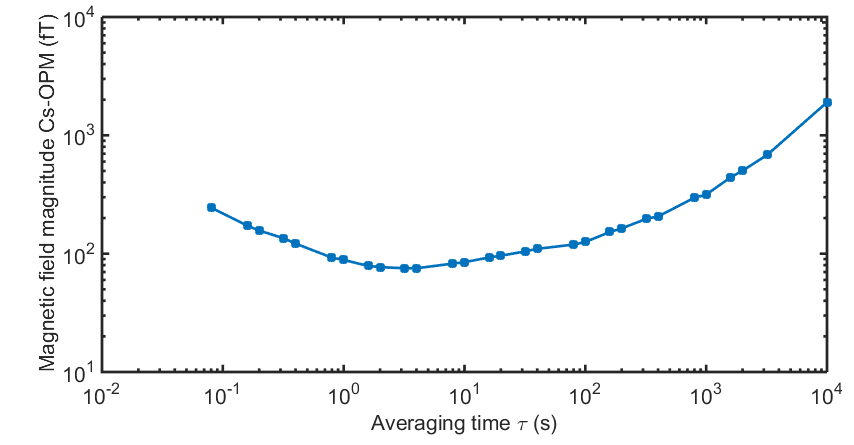}%
		\caption{Allan deviation of the magnitude of the magnetic field inside the four layer magnetic shield at PSI measured by a three-axis atomic vapor (Cs) vector magnetometer, data from Ref.\,\cite{Afach2015OExpress}. Note that the main magnetic field of the experiment is aligned with $z$ and has a magnitude of about \SI{1}{\muT}.}%
\label{fig:AD_CsM}%
\end{figure}

Figure\,\ref{fig:AD_CsM} shows the Allan deviation measured using a three axis vector magnetometer\,\cite{Afach2015OExpress} 
inside the PSI magnetic shield.
An improvement of approximate 3 orders of magnitude can be observed when compared to the AD of the outside field. The AD minimum of the field magnitude $\sigma_{\rm AD} \approx \SI{80}{fT}$ is at just below $\tau = \SI{100}{s}$.
Again the field stability would not be nearly as good as required even if the electric field polarity were to be changed after every Ramsey cycle ($1/f=\tau_{\rm AD} \approx \SI{300}{s}$) and the relative long charging times (\SI{200}{s}) of the electrodes were significantly reduced. To circumvent this problem all recent and future searches use one or more monitor magnetometer(s) to correct for magnetic-field drifts.

\subsection{Magnetic field monitors} 
As we have seen, fluctuating magnetic fields add noise to the measured nEDM data and limit the sensitivity of the experiment.
Using standard error propagation on equation\,\eqref{eq:DiffConfig2} one finds 

\begin{equation}
	\delta \dn = \sqrt{\sigma^2(\dn) + 2\left( \frac{\mu_{\rm n}}{2|E|}\delta B\right)^2}.
\label{eq:errorPropagation}
\end{equation}

\noindent Here the first term is the sensitivity due to neutron counting statistics, typically referred to
as the statistical sensitivity. The second term is the systematic effect that worsen
the counting statistics due to its stochastic nature. In general one would like to have the second term at least a factor of four smaller than the first one so that it would have only a
minor effect on the sensitivity. This can be achieved using additional magnetometers to compensate changes of the magnetic field between measurements. This requires a magnetometer with an accuracy of better than

\begin{equation}
	\delta B_\text{mag}  \leq \frac{\sigma(\dn)}{4} \frac{\left|E\right|}{\mu_\text{n}}
\label{eq:SensMagn}
\end{equation}

\noindent per Ramsey cycle. The proportionality constant for a $\unit[12]{kV/cm}$ field is $\mu_\text{n}/\left|E\right| = \unit[6\!\times\!10^{-27}]{\ecm/fT}$. Hence, the intrinsic sensitivity of the monitor magnetometer must be better than $\sigma(B_\text{mag})\leq \SI{5}{fT}$ for a statistical sensitivity of $\unit[\pow{1}{-25}]{\ecm}$ per cycle, a value typically aimed for by many new searches. Conceptually this might be accomplished by either using several high-sensitivity magnetometers, for example optically pumped atomic vapor magnetometers\,\cite{Budker2007} or SQUIDS\,\cite{Burghoff2007}, in the close vicinity of the neutron precession cell, or by using a magnetically susceptible isotope that can be injected into the same volume where the neutrons precess (cohabiting magnetometer).

\subsubsection{Arrays of high-precision magnetometers for field monitoring}

Point like (actually spheres with $r=\SI{15}{mm}$) magnetometers placed just outside of the precession chamber but still inside the magnetic shield are  often proposed or used to monitor the magnetic field.  They have the clear advantage not to see the electric field and hence should be free of an electric-field correlated effect. 

In the past, several experiments inferred the average magnetic field to which the neutrons were exposed by combining the values of several optically pumped atomic magnetometers (OPM)\,\cite{Smith1990PLB,Altarev1996,Serebrov2015PRC}. Further, the currently operating experiment at the PSI uses in addition to a cohabiting magnetometer an array of OPMs to extract magnetic field information. All of these magnetometers are scalar magnetometers, essentially measuring the magnitude of the magnetic field at a point in space\,\cite{Groeger2006}. By combining the measurement values of these OPM it is possible to extract a mean value and higher-order magnetic-field terms like gradients\,\cite{Afach2014PLB}. 

In the case of homogeneous field changes affecting all magnetometers in the same way, the mean value may be used to correct for magnetic field drifts. However, if a field change occurs due to a change in magnetization within or close to the observed volume, for example provoked by a leakage current across the insulator ring (see Sec.\,\ref{sec:correlatedSys} and references therein), the combined value may show no field change even though the neutrons are exposed to a different field. This could effectively lead to the observation of an EDM signal, as pointed out in Ref.\,\cite{Lamoreaux2009}. 

In Ref.\,\cite{Afach2014PLB} optical pump cesium magnetometers (Cs-OPM) were successfully used to decompose the magnetic field of the precession chamber into spherical polynomial harmonics to extract the vertical magnetic-field gradient $\partial B_z/\partial z$. New developments focusing on accurate vector information from OPM\,\cite{Afach2015OExpress} will further refine such methods and could prove useful in the future to identify and correct for systematic effects, as from a leakage current or by magnetic-field gradients $\partial B_z/\partial z$ in equations\,(\ref{eq:falseNeutronEDMHg}, \ref{eq:neutronGeomEffect}).

\subsubsection{Cohabiting magnetometers}
\begin{figure}%
	\includegraphics[width=0.9\columnwidth]{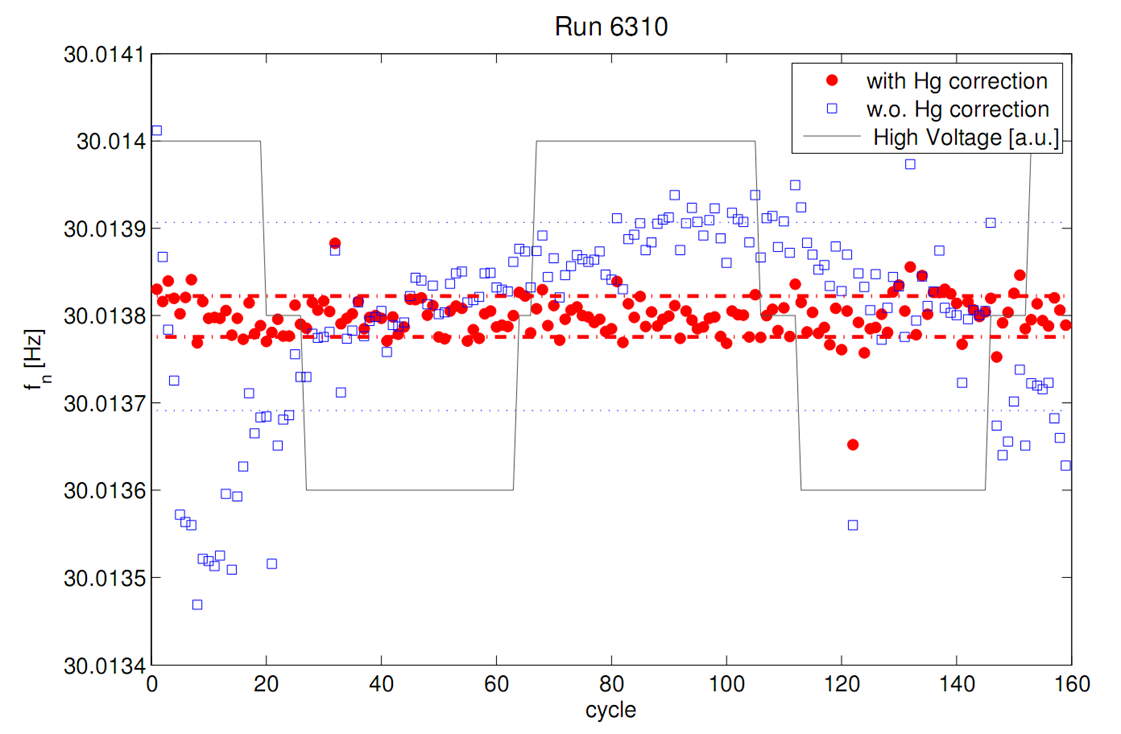}%
\caption{Neutron resonant frequency before and after correction for magnetic-field drifts using a mercury comagnetometer (data sample from nEDM-PSI collaboration\,\cite{Baker2011}.}%
\label{fig:HgCorrectednFrequency}%
\end{figure}

The second type of field monitor, using magnetically susceptible isotopes that occupy the same volume as the neutrons, is considered essential to most future experiments and is currently also in use within the spectrometer at PSI\@. Three different isotopes, all of nuclear spin 1/2, are typically considered for this task: \tHe\,\cite{Golub1983JPL,Ramsey1984}, \magHg{} and \magXe. While \magHg{} is already employed\,\cite{Green1998} the other two are proposed for spectrometers in the future. Although these magnetometers allow for a cycle-to-cycle correction of the neutron precession frequency as shown in Fig.\,\ref{fig:HgCorrectednFrequency}, they pose a certain risk of transferring systematic effects to the neutron measurement. Furthermore, even if all correlated systematic effects (Sec.\,\ref{sec:correlatedSys}) from a co-magnetometer are well under control a certain dilution of the neutron sensitivity may occur.

Ultracold neutrons sample the magnetic field of the precession cell slightly differently then the atoms of the thermal co-magnetometer gas. Stored ensembles of UCN have very low energies which leads to the awkward effect that their center-of-mass $\left\langle z_\text{\tiny UCN}\right\rangle$ is slightly below the geometrical center-of-mass $z_{\rm cm}$ of the precession cell. In the presence of a vertical magnetic-field gradient  $g_z=\partial B_z/\partial z$  this leads to an uncompensated contribution which adds to equation\,\eqref{eq:errorPropagation}:	

\begin{equation}
	\delta \dn = \sqrt{\sigma^2(\dn) + 2\left( \frac{\mu_{\rm n}}{2|E|}\delta B\right)^2 + 2\left(\frac{\mu_{\rm n}}{2|E|}\Delta h_{\rm cm} \delta g_z \right)^2},
\label{eq:errorPropagation2}
\end{equation}

\noindent with the center-of-mass offset $\Delta h_{\rm cm} = (z_{\rm cm}-z_{\rm UCN})\approx \SI{4}{mm}$\,\cite{Afach2015PRD,Afach2015PRL}, for the PSI spectrometer, depending on the UCN energy spectrum. If the random variation of the magnetic-field gradient between electric-field configuration is below 

\begin{equation}
	\delta g_z \leq \frac{\sigma(\dn)}{4\Delta h_{\rm cm}} \frac{\left|E\right|}{\mu_\text{n}},
\label{eq:GradientStabilityRequirement}
\end{equation}  

\noindent then no monitor for the gradient may be required: this is typically the case for E-field reversal every \SI{4.5}{\hour} at the PSI experiment. Otherwise, a gradiometer is needed to correct for changes with a sensitivity better than \SI{12}{\femto\tesla\per\centi\meter} for a neutron-only statistical sensitivity of \unit[$\pow{1}{-25}$]{\ecm} per cycle.

An even more subtle effect can further reduce the statistical sensitivity when using slowly oscillating magnetic fields to flip the co-magnetometer spin into the transverse plane. After the neutrons are locked into the cell the polarized co-magnetometer gas is injected and a $\pi/2$-pulse is applied to tip the spin of the co-magnetometer atoms by $\pi/2$. This circular rotating magnetic field is also seen by the neutrons and can lead to a small tilt of the neutron spin.
Using Rabi's equation for a resonant transition from $m = +1/2$ to $m=-1/2$\,\cite{Rabi1954}:

\begin{equation}
	 P_{-1/2} = \frac{\omega_1}{\Omega}\sin\left(\frac{\Omega t_{\pi/2}}{2}\right),
\label{eq:RabisEquation}
\end{equation}

\noindent with
\[
\omega_1 = \frac{\pi\gamma_n}{2\gamma_i t_{\pi/2}}~,\quad\Omega=\sqrt{\left(\omega_{\rm n} - \omega_{\pi/2}\right)^2-\omega_1^2}\quad\text{and}\quad\omega_{\pi/2}\approx \gamma_i B_0
\]
 it is possible to calculate this effect. This leads to tilting of the neutron spins by:

\begin{equation}	
 \Theta = \frac{R\pi}{\Omega t_{\pi/2}}\sin\left(\frac{\Omega t_{\pi/2}}{2}\right),
\label{eq:XeSpinTilt}
\end{equation}

\noindent where $R = \gamma_{\rm n}/\gamma_i$ is the ratio of gyromagnetic ratios of the neutron and the co-magnetometer isotope.

\begin{figure}%
\includegraphics[width=0.9\columnwidth]{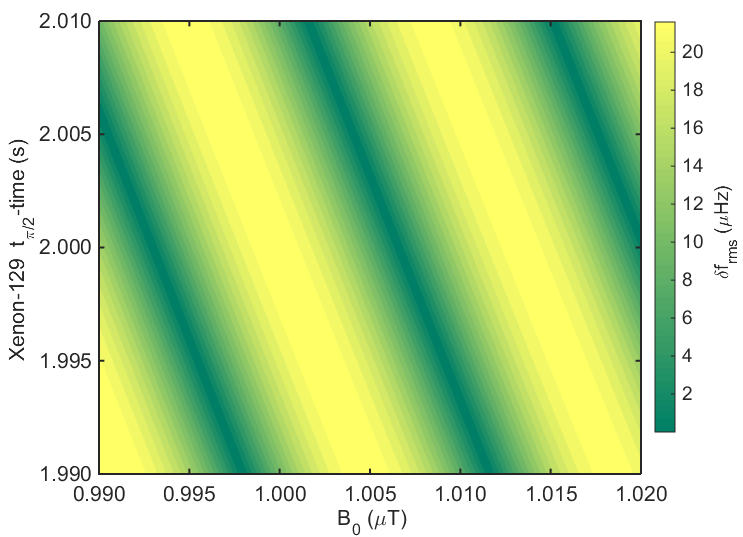}%
\caption{Systematic frequency error as a function of xenon $\pi/2$-pulse duration $t_{\pi/2}$ and main magnetic field $B_0$.}%
\label{fig:XeFrequencyShift}%
\end{figure}

If the phase relation and timing of the co-magnetometer $\pi/2$~pulse and the neutron $\pi/2$~pulse stays constant over time this only leads to a global shift of the neutron resonant frequency which cancels in the nEDM-analysis. However, if phase and timing is not controlled this can lead to a frequency uncertainty of

\begin{equation}
	\delta f_{\rm rms} = \frac{1}{\sqrt{2}}\frac{|\Theta|}{2\pi T},
\label{eq:FrequencyShiftPulse}
\end{equation}

\noindent where $T$ is the free precession time. Figure\,\ref{fig:XeFrequencyShift} shows the systematic uncertainty from this effect on the neutron precession frequency in the case of a \magXe\ co-magnetometer. For xenon this effect may be as large as \SI{20}{\micro\hertz} which would be approximately a factor twenty larger than the intrinsic sensitivity of the neutrons in a next generation experiment. Therefore it will be of paramount importance to adjust $t_{\pi/2}$ and $B_0$ such that $\sin\left(\Omega t_{\pi/2}/2\right)=0$ and to control phase and timing of the pulse generation. Furthermore it is possible to reduce this effect by choosing an appropriate envelope of the pulse, such as a triangular or a Hann window. 

\subsection{Effects mimicking an EDM signal}
\label{sec:correlatedSys}

Direct effects which mimic an EDM arise from a change of the magnetic field $\delta B(E)$ caused by the electric field. The strength of a false EDM signal can be approximated as in the previous section with $d_{\rm n, f} = \mu_{\rm n}/(2E)\left(\delta B (E^{\parallel}) -\delta B (E^{\nparallel})\right)$.
\subsubsection{Leakage currents}
\begin{figure}%
\includegraphics[width=0.7\columnwidth]{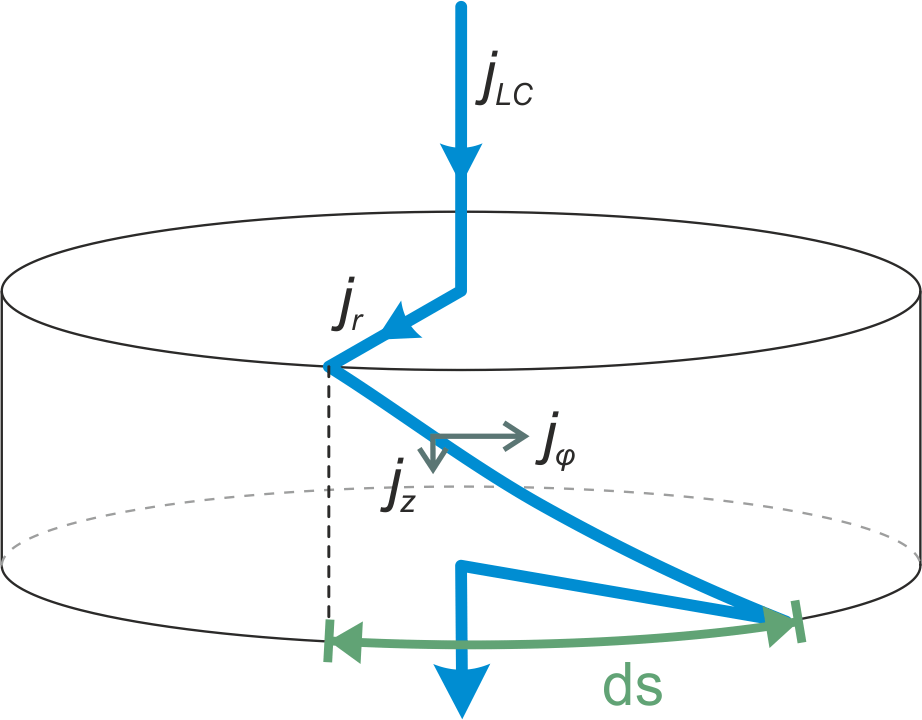}%
\caption{Idealized sketch of the path a leakage current may flow along. The currents flowing along the electrode in the radial direction $j_r$ produce the same magnetic field, in opposite direction, which cancel to first order. Further, the field generated by the fractional current along $z$ parallel to the cylinder axis $j_z$only generates a field in the transverse plane, which is strongly suppressed since the neutrons only see the magnitude of the field, which is dominated by $B_0$.} %
\label{fig:LeakageCurrentSketch}%
\end{figure}

Probably the most obvious direct effect is a leakage current from the charged electrode to ground along a given path. Such a path might result from material imperfections of the insulator ring or from discharges along the wall ( which can lead to visible traces, as reported by several experiments). A current flowing along this path will create a magnetic field which will be seen by neutrons. By reversing the high voltage the current will flow in the opposite direction and create a field with opposite sign, hence  $\left(\delta B (E^{\parallel}) -\delta B (E^{\nparallel})\right)= 2\delta B (E^{\parallel})$ if the current is of the same magnitude and follows the same path. Figure\,\ref{fig:LeakageCurrentSketch} shows the idealized geometry of the scenario discussed. Several different current paths were studied in Ref.\,\cite{Zenner2013} using finite elements. The most stringent limit is given by a path which makes a full circle around the insulator ring at the central plane. In this case the neutrons are exposed to a change of the magnetic field of $|\delta B (E^{\parallel})|=\SI{2}{\femto\tesla}$ for a current of \SI{1}{\nano\ampere}, which results in a false nEDM signal of $d_{\rm n,f} \approx \unit[\pow{1}{-26}]{\ecm}$. However, such a current path seems rather unlikely for which reason it became commonly accepted to estimate the effect by a current path as depicted in Fig.\,\ref{fig:LeakageCurrentSketch} with an azimuthal path length of $\diff s = \SI{0.1}{m}$, which means  the discharge path is running at up to about 45 degrees. In this case the false nEDM signal would only be $\unit[\pow{5}{-28}]{\ecm}$.
In general most spectrometer designs include a leakage-current monitor which records the integral current flowing across the insulator ring. Additional magnetometers, especially co-magnetometers, will also be sensitive to leakage-current induced field changes and may be used for correction. However, it has been reported that the mean value of scalar magnetometers placed in the close vicinity of the precession chamber might indicate no change in the magnetic field depending on their exact placement, despite the presence of a leakage current large enough to lead to a false nEDM signal. For a detailed discussion of this effect I refer the interested reader to Ref.\,\cite{Lamoreaux2009}.

\subsubsection{Systematic effects from electric fields}
Any particle moving through an electric field will see a motional magnetic field seen in its rest frame. According to special relativity the neutron is then exposed to an additional field of:

\begin{equation}
		B_{v\times E} = -\gamma\frac{\mbf{v}\times\mbf{E}}{c^2},
\label{eq:vxEEffect}
\end{equation}

\noindent where $\gamma = \left(1-v^2/c^2\right)^{-1/2}$ is approximately 1 for cold and ultracold neutrons.  This results in direct systematic effects which lead to a signal mimicking a true nEDM. 

For simplicity we consider the case where the electric field and the  magnetic field are slightly tilted by an angle $\theta$ with respect to each other in the $x$-$z$-plane ($\mbf{E} = \sin\!\theta\,E, 0 ,\cos\!\theta\,E$) and where a uniform magnetic field gradient of the form $\partial B_z/\partial z = -1/2(\partial B_x/\partial x +\partial B_y/\partial y)$ is present. In this case with $\cos\theta\approx 1$ the magnetic field in the rest frame of the particle moving through the cell at a point in space is 

\begin{align}
		B_x  =& \quad -\frac{x}{2}\frac{\partial B_z}{\partial z} -\frac{v_y E}{c^2} \nonumber \\
		B_y  =& \quad -\frac{y}{2}\frac{\partial B_z}{\partial z} -\frac{v_x E}{c^2}+\frac{\theta v_zE}{c^2} \\
		B_z  =& B_0 + z\frac{\partial B_z}{\partial z}+\frac{\theta v_yE}{c^2}. \nonumber
\label{eq:GeomPhase1}
\end{align}

\noindent This leads to a field magnitude of
\begin{align}
		|B(\mbf{r})| = B_0 &+ z\frac{\partial B_z}{\partial z}+\frac{x^2+y^2}{8B_0}\left(\frac{\partial B_z}{\partial z}\right)^2 \\		
		&+\frac{\theta v_yE}{c^2} \label{eq:linearvCrossETerm}\\
		&+\frac{\left(xv_y+yv_x+\theta y v_z\right)}{2B_0c^2}\frac{\partial B_z}{\partial z}E \label{eq:GeomTerm}\\
		&+\frac{v_y^2+\left(v_x-\theta v_z\right)^2}{2B_0c^4}E^2.
\label{eq:quadraticTermInE}
\end{align}

While the first line (equation 36) is free of any electric-field systematic the following three lines (eqs. 37-39) either have terms linear or quadratic in $E$. The effective magnitude of these terms depends on how the particle samples the precession cell and is discussed in the following paragraphs.
\medskip
\paragraph*{Motional magnetic field in the rest frame of a particle moving through an electric field}
\smallskip 
~\\
The latest beam experiment\,\cite{Dress1977} was limited by the motional magnetic field, the linear term in $E$ without gradient in equation\,\eqref{eq:linearvCrossETerm}, seen in the rest frame of a neutron moving through an electric field.
For the configuration of this last beam experiment, with $\mbf{v}= v_y = \unit[150]{m/s}$ and a field strength of \unit[100]{kV/cm}, the term $B_E = v_y E/c^2 =\SI{17}{nT}$.
In the case where the electric field and the magnetic field have a relative angle of $\theta$ the neutrons see a field change of $\Delta B(E) = 2\theta\cdot B_{E}$ when changing polarity of the electric field. The reported error of $\unit[\pow{1.5}{-24}]{\ecm}$ is equivalent to a misalignment angle of \SI{0.1}{\milli\radian}, which is extremely demanding. A proposal for a future beam experiment\,\cite{Piegsa2013PRC} intends to circumvent this issue by measuring the nEDM as a function of velocity and then extrapolating to $\mbf{v} = 0$.

Storage experiments using UCN are much less exposed to this effect, as the velocities are a factor 40 lower and in the limit of chaotic trajectories the mean value averages to zero. However, a possible effect which could lead to a relevant signal from $\mbf{v}\times \mbf{E}$ is a remaining ordered motion within the precession chamber.

Such a remaining common rotational motion could result from UCN which enter the precession chamber through an off-axis door or with a predominant tangential velocity component (twist). This could eventually lead to the situation that a fraction of the neutrons move on orbits along the insulator ring resulting in a net rotational flow with velocity $\mbf{v}_{\rm t}$. In this case one can replace $v_y$ with $v_r$ in equation\,\eqref{eq:linearvCrossETerm}, hence, $B_{\rm E}^{\rm rot} = \theta\frac{v_r\cdot E}{c^2}$ would be the result. By increasing the roughness of the precession chamber walls it is possible to reduce the probability of specular reflection which in turn would reduce the probability of ordered motion. Simulations\,\cite{Zenner2013} using Monte Carlo methods for particle trajectories show that an average $v_r < \unit[0.005]{m/s}$ can be obtained for a typical surface roughness. Together with a mechanically feasible alignment of the electric field relative to the magnetic field of better than $\theta \approx
 \SI{1}{\milli\radian}$ a remaining systematic effect of $d_{\rm n,f}\leq \unit[\pow{1.7}{-27}]{\ecm}$ seems likely. 
The detailed analysis of this effect by Pendlebury and coworkers\,\cite{Pendlebury2015PRD} give a limit of
$\unit[\pow{1}{-29}]{\ecm}$ arguing that by the time the first $\pi/2$-pulse is applied, tipping the neutrons into the transverse plane, the velocities of the stored UCN are sufficiently randomized.

\medskip
\paragraph*{Motional magnetic field from the quadratic term of $\mbf{v}\times E$}
\smallskip ~\\
Another effect correlated to the motion of particles in an electric field is the quadratic term\,\eqref{eq:quadraticTermInE}, which persists even if $\overline{\mbf{v}} =0$ is fulfilled. 
Simply taking the values for the mercury co-magnetometer\,\cite{Baker2014}, $B_0 = \SI{1}{\muT}$, $v_{\rm Hg}^2 = \left(\SI{150}{m/s}\right)^2$ and $E =\SI{1}{kV/cm}$ results in $ B_{\rm E^2} = \SI{1.4}{pT}$.
 This effect is transferred to the neutrons when using the mercury frequency as field monitor. Only a difference in magnitude of the E-field for both polarities will have a remaining contribution to a false nEDM signal. To keep this systematic effect below the current statistical sensitivity the electric-field magnitude would have to be equal to within one part in 10\,000. However, this estimate does not take into account the fact that the direction and velocity of the mercury atoms change with every collision. Hence, the atoms are effectively exposed to a fast and random oscillating motional magnetic field, and the field only has a constant magnitude for $\tau_c$, the average time between two collisions.

The effective magnetic-field shift for this situation is derived as a frequency shift for spin-1/2 particles in Ref.\,\cite{Lamoreaux2009} and can be written as

\begin{equation}
	B_{\rm E^2} = \frac{1}{2\gamma}\left\langle \left(B_{\rm v\times E}\right)^2 \frac{\gamma B_0\tau_c - \sin\left( \gamma B_0 \tau_c\right)}{\left(\gamma B_0\right)^2  \tau_c} \right\rangle_{\rm av}.
\label{eq:vCrossECubicShift}
\end{equation}

\noindent The index `av' indicates an average over the statistical ensemble and for times much longer than $\tau_c$. In a nEDM experiment with a co-magnetometer two extreme cases exist. For $\gamma B_0\tau_c \gg 1$ the term $\sin\left( \gamma B_0 \tau_c\right)$ has a zero ensemble average. Further the average over all possible direction for $\mbf{v}$ gives a factor 2/3. Therefore the quadratic motional magnetic field is

\begin{equation}
	B_{\rm E^2} = 2/3\frac{v^2 E^2}{2B_0c^4}.
\label{eq:vCrossECubicUCN}
\end{equation}

\noindent In measurements using UCN we are in this regime. For an average velocity of $v = \SI{3}{m/s}$, an electric field of \SI{12}{kV/cm}, $B_0=\SI{1}{\micro\tesla}$ we have a motional field of  $B_{\rm E^2}^{\rm UCN} =\SI{0.5}{fT}$. This is completely negligible as the effect only shows up if the magnitude for positive and negative electric fields are different. Modern high voltage power supplies have specifications of better than $|U^+|-|U^-| < 0.001 U$ with $U$ the nominal voltage, which further suppresses this systematic effect by a factor thousand, resulting in a limit of $d_{\rm n,E^2} < \unit[\pow{3}{-30}]{\ecm}$. 

In the case when  $\gamma B_0\tau_c \ll 1$ the sine-term in equation\,\eqref{eq:vCrossECubicShift} can be expanded:

\begin{equation}
	B_{\rm E^2}=2/3\frac{v^2 E^2}{2B_0c^4} \frac{\left(\gamma B_0 \tau_c\right)^2}{6},
\label{eq:vCrossECubicHg}
\end{equation}

\noindent with $\tau_c \approx \unit[\pow{1}{-3}]{s}$ for a precession cell of radius $r=\SI{0.25}{m}$ and height $h=\SI{0.12}{m}$. Mercury atoms in the ballistic limit with $\gamma = \gamma_{\rm Hg}$ match this case very well. For the above parameters this gives a motional magnetic field of $B_{\rm E^2}^{\rm Hg} \leq \SI{5}{\atto\tesla}$ due to the large suppression factor of $\pow{1}{-4}$ of the last quadratic term.
\medskip

\paragraph*{Frequency shifts of particles moving in an inhomogenous magnetic field and an electric field} \smallskip
~\\
Currently the most important correlated systematic effect results from the term in equation\,\eqref{eq:GeomTerm} linear in gradient and electric field. It is of the same order of magnitude as the statistical sensitivity\,\cite{Pignol2012PRA}. 
It gives rise to a frequency shift which depends on the way that the particle is sampling the field of the precession chamber. This effect was first discussed in the limit of a trumped shaped field with a vertical gradient $\partial B_z/\partial z = -1/2(\partial B_x/\partial x +\partial B_y/\partial y $ and an electric field\,\cite{Pendlebury2004}.  An approach using the general theory of relaxation with correlation functions was then used in Ref.\,\cite{Lamoreaux2005,Barabanov2006} but still applied to a cylindrical uniform field gradient. In the article of Pignol and Roccia\,\cite{Pignol2012PRA} a general form for arbitrary fields and geometries is given:

\begin{equation}
	\delta \omega = -\frac{\gamma^2 E}{c}\left\langle xB_x + yB_y\right\rangle,
\label{eq:GenGeomShift}
\end{equation}

\noindent where the brackets refer to a volume average over the cell. For the further discussion we will use the expressions from Ref.\,\cite{Afach2015EPJD} which has confirmed the theoretical predictions in a direct measurement with an array of scalar OPM around the central precession cell of a mercury magnetometer. A UCN based experiment using a thermal co-magnetometer works in the regime of two limiting cases: 

\begin{align}
	\delta \omega= \frac{\gamma D^2}{16c^2}\frac{\partial B_z}{\partial z}E \quad\quad & \text{non-adiabatic} \\
	\delta \omega = \frac{v_{xy}^2E}{2B_0c^2}\frac{\partial B_z}{\partial z} \quad\quad & \text{adiabatic}.
\label{eq:GeomPhase3}
\end{align}

\noindent It is worthwhile to note that the adiabatic case is independent of the gyromagnetic ratio of the particle and is known as geometric phase or Berry phase\,\cite{Berry1984,Commins1991}. Measurements using UCN fall into this regime. Using the relation $\dn = \hbar/2E\cdot\delta\omega$ the shift can be misinterpreted as an EDM of

\begin{equation}
	\dn^{\rm false} = \frac{\partial B_z}{\partial z}\cdot \unit[\pow{1.7}{-28}]{\ecm \frac{cm}{pT}},
\label{eq:neutronGeomEffect}
\end{equation}

\noindent for $v_xy = \SI{3}{m/s}$ and $B_0 = \SI{1}{\micro\tesla}$. For current experimental sensitivities this is not a problem, however, in future experiments aiming at sensitivities in the low $\unit[\pow{1}{-28}]{\ecm}$ region a perfect control of the magnetic field homogeneity is mandatory. In the case of \magHg\ the atoms sense the field in the non-adiabatic limit, and hence the false edm signal is

\begin{equation}
	d_{\rm Hg}^{\rm false} = \frac{\partial B_z}{\partial z}\cdot  \unit[\pow{1.15}{-27}]{\ecm  \frac{cm}{pT}},
\label{eq:falseEDMHg}
\end{equation}

\noindent which is transferred to the neutron measurement:

\begin{equation}
	d_{\rm Hg\rightarrow n}^{\rm false} =-\frac{\partial B_z}{\partial z} \cdot \unit[\pow{4.4}{-27}]{\ecm \frac{cm}{pT}}.
\label{eq:falseNeutronEDMHg}
\end{equation}

\begin{figure}%
	\includegraphics[width=0.95\columnwidth]{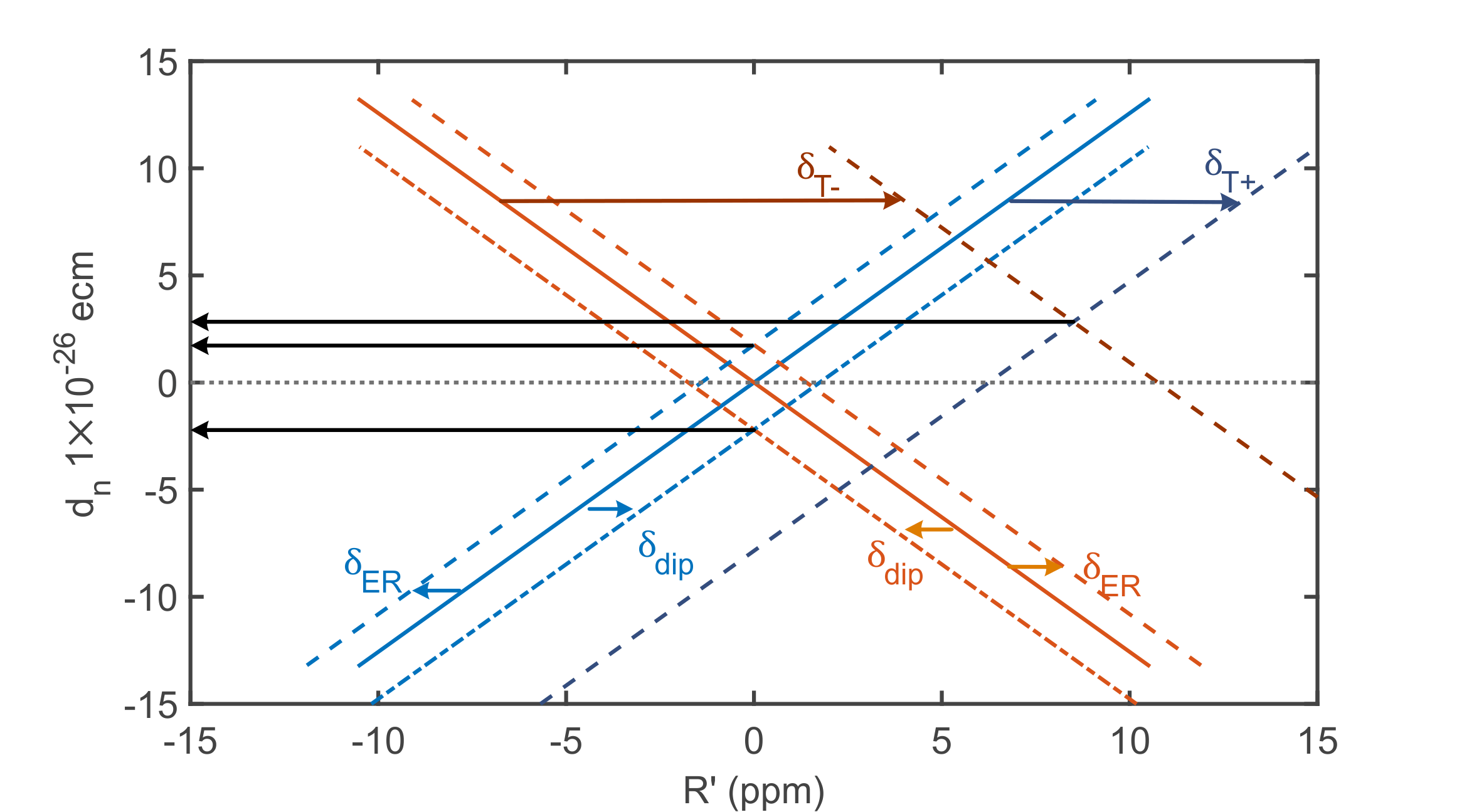}%
\caption{Neutron EDM as function of $R'$ for mercury as co-magnetometer, indicating systematic effects which may occur when using the crossing point analysis as in Ref.\,\cite{Pendlebury2015PRD}. The solid orange ($B_0^{\uparrow}$) and blue ($B_0^{\downarrow}$) line indicate the perfect systematic free measurement. Some possible shifts are indicated individually. Earth rotation shift $\delta_{\rm ER}^{\uparrow/\downarrow}$ for main magnetic field up/down; shift from a permanent dipole close to the precession chamber $\delta_{\rm dip}$; shifts from transverse magnetic fields of different magnitude $\delta_{\rm T}$. The black arrows pointing to the $y$-scale indicate the magnitude of the systematic effect. Note that, if the transverse fields for both magnetic field configurations is identical, $B_{\rm T}^{\uparrow} = B_{\rm T}^{\downarrow}$, the systematic effect vanishes.}%
\label{fig:GeomSysEffects}%
\end{figure}

Already gradients of some \pT/cm will be enough to fully dominate an otherwise very sensitive search for an nEDM\@. Ideally one would like to have a magnetometer scheme which is capable of measuring the gradient with a precision of at sub \pT/cm-resolution to correct for this effect. The current version of the cesium OPM array in the PSI experiment achieves an accuracy of approximately \unit[5-10]{\pT/cm} which is not sufficient for this purpose. An alternative approach is to bring the cohabiting magnetometer also into the adiabatic regime by using a sufficiently high pressure of buffer gas\,\cite{Masuda2012PLA}.
Currently the transferred effect can be mitigated by measuring \dn\ as a function of 

\begin{equation}
	R = \frac{\omega_{\rm n}}{\omega_{\rm Hg}}=\frac{\gamma_{\rm n}}{\gamma_{\rm Hg}}\left(1+\frac{g_z\!\cdot\!\Delta h_{\rm cm}}{B_0}+ \delta_{\rm T} + \delta_{\rm ER} + \delta_{\rm dip}+	\delta_{\rm LS}\right),
\label{eq:R}
\end{equation}

\noindent where $\Delta h_{\rm cm}$ is the center-of-mass offset between the UCN and the mercury gas. This is one example of the subtle differences in how UCN and thermal atoms sample the precession volume which lead to tiny magnetic-field dependent deviation of the frequencies of the neutron and the co-habiting isotope (for \magHg{} see Ref.\,\cite{Afach2014PLB}). Neglecting for the moment potential systematic effects, indicated with $\delta_{\dots}$ in equation\,\eqref{eq:R}, one can redefine a value

\begin{equation}
	R'-1 = R\frac{\gamma_{\rm Hg}}{\gamma_{\rm n}} -1 = \frac{g_z\!\cdot\!\Delta h_{\rm cm}}{B_0},
\label{eq:RPrimeMinusOne}
\end{equation}

which can be used as measurement of the vertical magnetic-field gradient $g_z$. The true value of the measured nEDM is the crossing point of two curves $\dn(R'-1)$ vs $(R'-1)$  shown in Fig.\,\ref{fig:GeomSysEffects} found by reversing $B_0$. However, a whole class of systematic effects appear, as $R'-1$ is not only sensitive to the vertical magnetic-field gradient, but also to all other effects which might shift the ratio of precession frequencies $R$.

The systematic shift from transverse fields is a result from the fact that the co-magnetometer operates in the non-adiabatic regime and essentially measures only the vectorial average of the magnetic field $|\langle \mbf{B} \rangle|$. Whereas the neutrons operate in the adiabatic regime and measure the volume averaged field modulus $\langle |\mbf{B}|\rangle$. Hence, in the presence of a residual transverse field like a quadrupole field perpendicular to $B_z$ with $B_x= qy$, $B_y= qx$ a frequency shift of

\begin{equation}
		\delta_T = \frac{\langle B_T^2\rangle}{2B_0^2}
\label{eq:transveresFieldSys}
\end{equation}

\noindent may occur, where $\langle B_T^2\rangle$ is the volume average over the transverse magnetic-field modulus squared. The systematic false effect from transverse fields is proportional to $\Delta B_{\rm T}^2 = \langle  B_{\rm T}^{\uparrow 2}\rangle -\langle B_{\rm T}^{\uparrow 2}\rangle$. A $\Delta B_{\rm T}^2  = \SI{1}{nT^2}$ shifts the crossing point EDM value by approximately $\unit[\pow{0.3}{-27}]{\ecm}$. It is possible to measure the transverse fields inside the precession chamber using a three-axis fluxgate to a precision of better than \SI{1}{nT}\,\cite{Afach2014PLB} which seems to be sufficient for current searches. In the future similar magnetic field maps recorded with a three-axis OPM\,\cite{Afach2015OExpress} might achieve a much better precision to permit searches in the low $\unit[\pow{1}{-27}]{\ecm}$ range.

The systematic effect $\delta_{\rm ER}$ is a shift due to Earth rotation which only appears if the gyromagnetic ratio of the neutron and the co-magnetometer isotope have different relative signs. The effective frequency shift depends on the latitude $\lambda$ of the location of the experiment on earth:

\begin{equation}
	\delta_{\rm ER}^{\uparrow/\downarrow} = -{\rm sgn}(B_0)\left(\frac{f_{\rm E}}{f_{\rm n}}+\frac{f_{\rm E}}{f_{\rm Hg}}\right)\sin\left(\lambda\right).
\label{eq:EarthRotation}
\end{equation}
 
Where the Earth's rotation frequency $f_{\rm E}=\SI{11.6}{\micro\hertz}$.
For the experiment at the Paul Scherrer Institute ($\lambda \approx \SI{46.6}{\degree}$ north) this results in a shift of $\delta_{\rm ER}^{\uparrow/\downarrow} =\mp \unit[\pow{1.4}{-6}]{Hz}$. 
Which, when forgotten, changes the crossing point to $d_{\times} = \unit[\pow{1.76}{-25}]{\ecm}$.

The next possible contribution comes from small magnetic permanent dipole fields close to the cell's surface. This will locally cause an inhomogeneous $\partial B_z /\partial z$ gradient which adds to $d_{\rm Hg\rightarrow n}^{\rm false}$ of equation\,\eqref{eq:falseNeutronEDMHg}. Further it will induce a frequency shift $\delta_{\rm dip}$ moving the lines horizontally by an equal amount but in opposite directions. Both the exact magnitude and orientation as well as position of a permanent dipole in the cell container influence the strength of the systematic effect. A possible method to avoid such an effect is to thoroughly scan all parts which constitute the precession cell assembly. Pendlebury and coworkers\,\cite{Pendlebury2015PRD} allowed for a systematic effect in their analysis of $\unit[\pow{\pm 6}{-27}]{\ecm}$ for an undetected permanent dipole moment in the mercury door.

\section{WORLD-WIDE EFFORTS} 
\label{sec:WWnEDM}
\begin{figure}%
\centering
  \includegraphics[width=0.7\columnwidth]{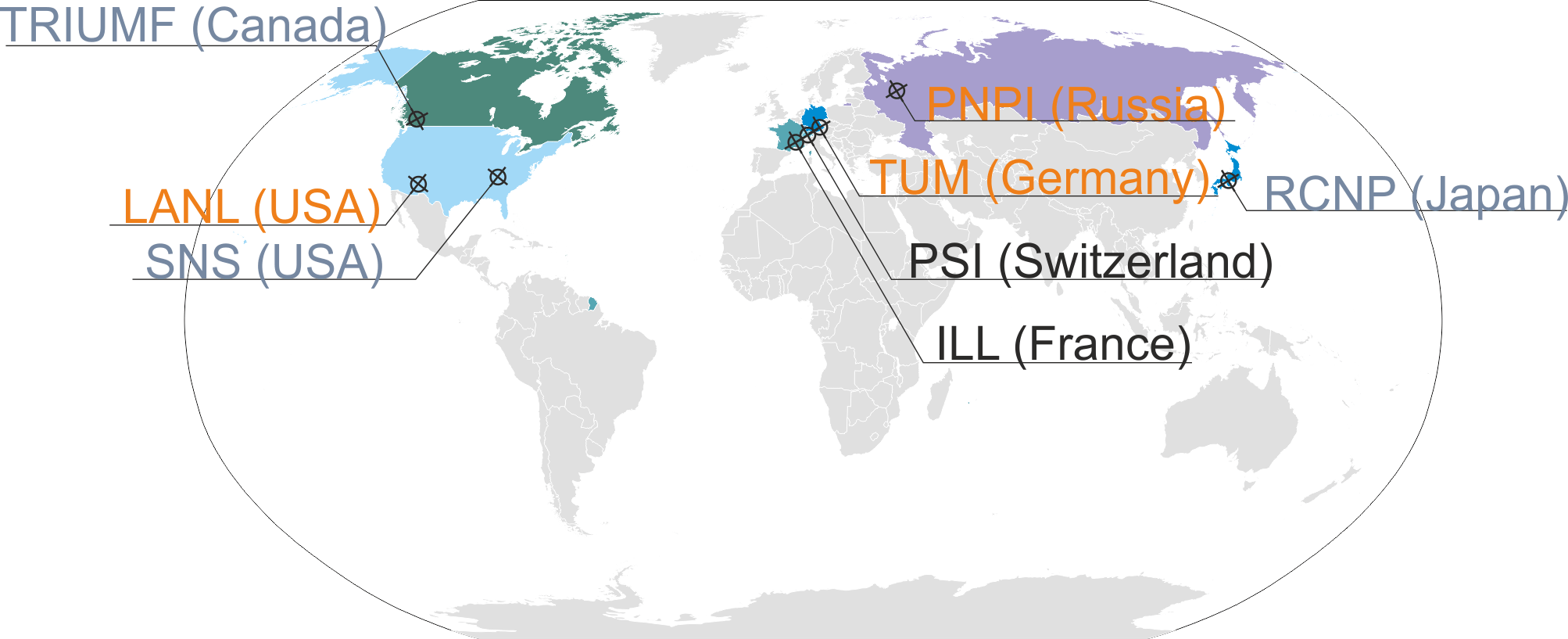}%
\caption{World map of nEDM searches. Black: UCN source and nEDM operational (ILL, PSI). Orange: Either nEDM spectrometer or UCN source operational. Blue: nEDM spectrometer and source in preparation.}%
\label{fig:WMapnEDM}%
\end{figure}

Several groups word wide, see Fig.\,\ref{fig:WMapnEDM}, compete in the effort to search for an electric dipole moment of the neutron. A detailed  summary of the individual projects can be found in Tab.\,\ref{tab:WWSearches} and in the references therein.

Currently only the collaboration at PSI is in the situation of being able to take data, aiming in the next year for a modest improvement of the current sensitivity into the low $\unit[10^{-26}]{\ecm}$ range. The PNPI group has upgraded its spectrometer and plans to continue data-taking at ILL in the next years.
Both efforts are essentially limited by UCN counting statistics, for which reason the groups already work on or plan upgrades either by constructing a new spectrometer which is better adapted to the UCN source (PSI), or by moving to a still-to-be-constructed better source (ILL $\rightarrow$ PNPI).
All next generation designs aim for a sensitivity in the low $10^{-27}$ or even in the $\unit[10^{-28}]{\ecm}$ range. The concepts rely on new UCN sources based on superthermal conversion, either using superfluid helium or solid deuterium, promising to deliver at least two orders of magnitude more UCN\@. 

Two other projects stand out as they are not using UCN, but cold neutrons and their distinct techniques have not been discussed in these proceedings. The crystal diffraction nEDM experiment at ILL, which is currently being upgraded, uses the very high electric field of crystals without center of symmetry\,\cite{Fedorov2011PhyB}. Finally, the project in discussion for the European Spallations Source at Lund is proposing a double beam experiment. The pulsed beam of the spallation source lends itself ideally to a measurement of the $\mathbf{v}\times\mathbf{E}$-effect as a function of velocity, from which an extrapolation to zero velocity would yield the real nEDM value\,\cite{Piegsa2013PRC}. 

\begin{table}%
\centering
\begin{tabular}{lcrcr}
\hline
	\bf Facility,  &\bf UCN (CN) source &\bf Spectrometer &\bf Magnetometry  &\bf Sensitivity \\
	 location&  &  &  &  $\mathcal{O}(\unit[10^{-27}]{\ecm}$)\\
	\hline
	LANL, USA & sD$_2$\,\cite{Saunders2013RSI} & dbl. chamber & \magHg  & 1 \\
	ILL, France & cold source\cite{Ageron1978},  & dbl.\ chamber\,\cite{Serebrov2015PRC} & Cs-OPM  & 10 \\
							&turbine\,\cite{Steyerl1986PLA}& & &  \\
  ILL, France & He-II to vacuum\,\cite{Zimmer2015PRC} & dbl. chamber\,\cite{Altarev2012a} & \magHg, Cs-OPM  & 1\\
	ILL, France & CN beam\,\cite{Abele2006} & crystal-diffraction\,\cite{Fedorov2011PhyB} & -  & 10 \\
	ESS, Sweden & pulsed CN & dbl.\ chamber\,\cite{Piegsa2013PRC}& - & 0.1 -10 \\
	FRM-II, Germany & sD$_2$\,\cite{Trinks2000NIMA} & dbl.\ chamber\,\cite{Altarev2012a}& \magHg, Cs-OPM  & 0.1 \\
	SNS, Oakridge & He-II in-situ\,\cite{Kolarkar2010AIP} & dbl.\ chamber\,\cite{Kolarkar2010AIP} & \tHe{}, squids & <0.1 \\
	PNPI, Russia& He-II to vacuum\,\cite{Serebrov2014TPL} & dbl.\ chamber\,\cite{Serebrov2015PRC} & Cs-OPM  & 0.1-1 \\ 
	TRIUMF, Canada & He-II to vacuum\,\cite{Masuda2015JPS} & dbl.\ chamber\,\cite{Matsuta2013AIP} & \magHg, \magXe{} & 0.1-1 \\
	PSI, Switzerland & sD$_2$\,\cite{Lauss2014} & sgl./dbl.\ chamber\,\cite{Baker2011}& \magHg, \tHe, Cs-OPM  & 1-10\\
	
\end{tabular}
\caption{Overview of all current nEDM-projects. Several groups plan to start data-taking in the next years to improve the current sensitivity by a factor ten. Another order of magnitude improvement seems possible until the middle of the next decade.}
\label{tab:WWSearches}
\end{table}

%
%

\section{CONCLUSION}
Several groups worldwide are pursuing promising approaches to improve the sensitivity for a permanent electric dipole moment of a fundamental particle. In the next few years a new result will probably improve the current limit on the nEDM slightly, while an order of magnitude improvement seems likely within the next five years or so. Another order of magnitude will require a breakthrough in UCN counting statistics or a dramatic improvement of the electric field which seems to be possible in superfluid helium.


\section{ACKNOWLEDGMENTS}
I would like to thank all my colleagues from the UCN group at PSI and the the nEDM-collaboration for many discussions. In particular I would like to thank 
N.~Ayers, K.~Kirch and P.~Harris for reading the manuscript and providing important suggestions and advice.
%



\begin{thebibliography}{100}
\providecommand{\url}[1]{\texttt{#1}}
\providecommand{\urlprefix}{URL }

\bibitem{Engelkemeir1962PR}
D.~W. Engelkemeir, K.~F. Flynn, \& L.~E. Glendenin.
\newblock Phys. Rev. \textbf{126} (1962) 1818.
\newblock \href {http://dx.doi.org/10.1103/PhysRev.126.1818}
  {\path{doi:10.1103/PhysRev.126.1818}}.

\bibitem{Riotto1999ARNPS}
A.~{Riotto} \& M.~{Trodden}.
\newblock Annu. Rev. Nucl. Part. Sci. \textbf{49} (1999) 35.
\newblock \href {http://arxiv.org/abs/hep-ph/9901362}
  {\path{arXiv:hep-ph/9901362}}, \href
  {http://dx.doi.org/10.1146/annurev.nucl.49.1.35}
  {\path{doi:10.1146/annurev.nucl.49.1.35}}.

\bibitem{Morrissey2012NJP}
D.~E. Morrissey \& M.~J. Ramsey-Musolf.
\newblock New J. Phys. \textbf{14} (2012) 125003.
\newblock \href {http://arxiv.org/abs/1206.2942} {\path{arXiv:1206.2942}},
  \href {http://dx.doi.org/10.1088/1367-2630/14/12/125003}
  {\path{doi:10.1088/1367-2630/14/12/125003}}.

\bibitem{Dine2012}
M.~Dine \& A.~Kusenko.
\newblock Rev. Mod. Phys. \textbf{76} (2003) 1.
\newblock \href {http://dx.doi.org/10.1103/RevModPhys.76.1}
  {\path{doi:10.1103/RevModPhys.76.1}}.

\bibitem{Cyburt2008JCAP}
R.~H. {Cyburt}, B.~D. {Fields}, \& K.~A. {Olive}.
\newblock \jcap \textbf{11} 012.
\newblock \href {http://arxiv.org/abs/0808.2818} {\path{arXiv:0808.2818}},
  \href {http://dx.doi.org/10.1088/1475-7516/2008/11/012}
  {\path{doi:10.1088/1475-7516/2008/11/012}}.

\bibitem{Aad2012PLB}
G.~Aad, T.~Abajyan, B.~Abbott, et~al.
\newblock \plb \textbf{716} (2012) 1 .
\newblock \href {http://dx.doi.org/10.1016/j.physletb.2012.08.020}
  {\path{doi:10.1016/j.physletb.2012.08.020}}.

\bibitem{Chatrchyan2012PLB}
S.~Chatrchyan, V.~Khachatryan, A.~Sirunyan, et~al.
\newblock \plb \textbf{716} (2012) 30 .
\newblock \href {http://dx.doi.org/10.1016/j.physletb.2012.08.021}
  {\path{doi:10.1016/j.physletb.2012.08.021}}.

\bibitem{Raidal2008}
M.~Raidal, A.~Schaaf, I.~Bigi, et~al.
\newblock \epjc \textbf{57} (2008) 13.
\newblock \href {http://arxiv.org/abs/0801.1826} {\path{arXiv:0801.1826}},
  \href {http://dx.doi.org/10.1140/epjc/s10052-008-0715-2}
  {\path{doi:10.1140/epjc/s10052-008-0715-2}}.

\bibitem{Sakharov1991}
A.~D. {Sakharov}.
\newblock \spu \textbf{34} (1991) 392.
\newblock \href {http://dx.doi.org/10.1070/PU1991v034n05ABEH002497}
  {\path{doi:10.1070/PU1991v034n05ABEH002497}}.

\bibitem{Baron2014Science}
J.~Baron, W.~C. Campbell, D.~DeMille, et~al.
\newblock Science \textbf{343} (2014) 269.
\newblock \href {http://dx.doi.org/10.1126/science.1248213}
  {\path{doi:10.1126/science.1248213}}.

\bibitem{Graner2016PRL}
B.~Graner, Y.~Chen, E.~G. Lindahl, \& B.~R. Heckel.
\newblock Phys. Rev. Lett. \textbf{116} (2016) 161601.
\newblock \href {http://arxiv.org/abs/1601.04339} {\path{arXiv:1601.04339}},
  \href {http://dx.doi.org/10.1103/PhysRevLett.116.161601}
  {\path{doi:10.1103/PhysRevLett.116.161601}}.

\bibitem{Pendlebury2015PRD}
J.~M. {Pendlebury}, S.~{Afach}, N.~J. {Ayres}, et~al.
\newblock \prd \textbf{92} 092003.
\newblock \href {http://arxiv.org/abs/1509.04411} {\path{arXiv:1509.04411}},
  \href {http://dx.doi.org/10.1103/PhysRevD.92.092003}
  {\path{doi:10.1103/PhysRevD.92.092003}}.

\bibitem{Baker2006}
C.~A. Baker, D.~D. Doyle, P.~Geltenbort, et~al.
\newblock Phys. Rev. Lett. \textbf{97} (2006) 131801.
\newblock \href {http://dx.doi.org/10.1103/PhysRevLett.97.131801}
  {\path{doi:10.1103/PhysRevLett.97.131801}}.

\bibitem{Pospelov2005}
M.~{Pospelov} \& A.~{Ritz}.
\newblock Ann. Phys. \textbf{318} (2005) 119.
\newblock \href {http://arxiv.org/abs/hep-ph/0504231}
  {\path{arXiv:hep-ph/0504231}}, \href
  {http://dx.doi.org/10.1016/j.aop.2005.04.002}
  {\path{doi:10.1016/j.aop.2005.04.002}}.

\bibitem{Engel2013PPNP}
J.~Engel, M.~J. Ramsey-Musolf, \& U.~van Kolck.
\newblock \ppnp \textbf{71} (2013) 21 .
\newblock \href
  {http://dx.doi.org/http://dx.doi.org/10.1016/j.ppnp.2013.03.003}
  {\path{doi:http://dx.doi.org/10.1016/j.ppnp.2013.03.003}}.

\bibitem{Chupp2015PR}
T.~Chupp \& M.~Ramsey-Musolf.
\newblock \prc \textbf{91} (2015) 035502.
\newblock \href {http://arxiv.org/abs/1407.1064} {\path{arXiv:1407.1064}},
  \href {http://dx.doi.org/10.1103/PhysRevC.91.035502}
  {\path{doi:10.1103/PhysRevC.91.035502}}.

\bibitem{Purcell1950PR}
E.~M. Purcell \& N.~F. Ramsey.
\newblock Phys. Rev. \textbf{78} (1950) 807.
\newblock \href {http://dx.doi.org/10.1103/PhysRev.78.807}
  {\path{doi:10.1103/PhysRev.78.807}}.

\bibitem{Smith1957PR}
J.~H. {Smith}, E.~M. {Purcell}, \& N.~F. {Ramsey}.
\newblock Phys. Rev. \textbf{108} (1957) 120.
\newblock \href {http://dx.doi.org/10.1103/PhysRev.108.120}
  {\path{doi:10.1103/PhysRev.108.120}}.

\bibitem{Dress1977}
W.~B. {Dress}, P.~D. {Miller}, J.~M. {Pendlebury}, et~al.
\newblock \prd \textbf{15} (1977) 9.
\newblock \href {http://dx.doi.org/10.1103/PhysRevD.15.9}
  {\path{doi:10.1103/PhysRevD.15.9}}.

\bibitem{Lushchikov1969JETPL}
V.~I. {Lushchikov}, Y.~N. {Pokotilovskii}, A.~V. {Strelkov}, \& F.~L.
  {Shapiro}.
\newblock J. Exp. Theor. Phys. \textbf{9} (1969) 23.
\newblock \urlprefix\url{http://adsabs.harvard.edu/abs/1969JETPL...9...23L}.

\bibitem{Shapiro1970SPU}
F.~Shapiro.
\newblock \spu \textbf{12} (1970) 580.
\newblock \href {http://dx.doi.org/10.1070/PU1970v012n04ABEH003909}
  {\path{doi:10.1070/PU1970v012n04ABEH003909}}.

\bibitem{Altarev1980NuPhA}
I.~S. {Altarev}, Y.~V. {Borisov}, A.~B. {Brandin}, et~al.
\newblock \nphysa \textbf{341} (1980) 269.
\newblock \href {http://dx.doi.org/10.1016/0375-9474(80)90313-9}
  {\path{doi:10.1016/0375-9474(80)90313-9}}.

\bibitem{Schmidt-Wellenburg2016LASNPA}
P.~{Schmidt-Wellenburg}.
\newblock ArXiv e-prints \href {http://arxiv.org/abs/1602.01997}
  {\path{arXiv:1602.01997}}.

\bibitem{Noether1918}
E.~Noether.
\newblock Nachrichten von der Gesellschaft der Wissenschaften zu G\"{o}ttingen,
  Mathematisch-Physikalische Klasse \textbf{1918} (1918) 235.

\bibitem{Christenson1964}
J.~H. Christenson, J.~W. Cronin, V.~L. Fitch, \& R.~Turlay.
\newblock Phys. Rev. Lett. \textbf{13} (1964) 138.
\newblock \href {http://dx.doi.org/10.1103/PhysRevLett.13.138}
  {\path{doi:10.1103/PhysRevLett.13.138}}.

\bibitem{Abe2001PRL}
K.~{Abe}, K.~{Abe}, R.~{Abe}, et~al.
\newblock \prl \textbf{87} (2001) 091802.
\newblock \href {http://arxiv.org/abs/hep-ex/0107061}
  {\path{arXiv:hep-ex/0107061}}, \href
  {http://dx.doi.org/10.1103/PhysRevLett.87.091802}
  {\path{doi:10.1103/PhysRevLett.87.091802}}.

\bibitem{Aubert2001}
B.~{Aubert}, D.~{Boutigny}, J.-M. {Gaillard}, et~al.
\newblock Phys. Rev. Lett. \textbf{87} (2001) 091801.
\newblock \href {http://arxiv.org/abs/hep-ex/0107013}
  {\path{arXiv:hep-ex/0107013}}, \href
  {http://dx.doi.org/10.1103/PhysRevLett.87.091801}
  {\path{doi:10.1103/PhysRevLett.87.091801}}.

\bibitem{Dubbers2011}
D.~{Dubbers} \& M.~G. {Schmidt}.
\newblock \rmp \textbf{83} (2011) 1111.
\newblock \href {http://arxiv.org/abs/1105.3694} {\path{arXiv:1105.3694}},
  \href {http://dx.doi.org/10.1103/RevModPhys.83.1111}
  {\path{doi:10.1103/RevModPhys.83.1111}}.

\bibitem{Greene1978}
G.~Greene.
\newblock \pra \textbf{18} (1978) 1057.
\newblock \href {http://dx.doi.org/10.1103/PhysRevA.18.1057}
  {\path{doi:10.1103/PhysRevA.18.1057}}.

\bibitem{Maggiore2005Book}
M.~Maggiore.
\newblock \emph{A Modern Introduction of Quantum Field Theory}.
\newblock Oxford University Press (2005).

\bibitem{Seng2015PRC}
C.-Y. Seng.
\newblock Phys. Rev. \textbf{C91} (2015) 025502.
\newblock \href {http://arxiv.org/abs/1411.1476} {\path{arXiv:1411.1476}},
  \href {http://dx.doi.org/10.1103/PhysRevC.91.025502}
  {\path{doi:10.1103/PhysRevC.91.025502}}.

\bibitem{Shabalin1983}
E.~Shabalin.
\newblock \spu \textbf{26} (1983) 297.
\newblock \href {http://dx.doi.org/10.1070/PU1983v026n04ABEH004331}
  {\path{doi:10.1070/PU1983v026n04ABEH004331}}.

\bibitem{Khriplovich1982PL}
I.~B. Khriplovich \& A.~R. Zhitnitsky.
\newblock \plb \textbf{109} (1982) 490.
\newblock \href {http://dx.doi.org/10.1016/0370-2693(82)91121-2}
  {\path{doi:10.1016/0370-2693(82)91121-2}}.

\bibitem{Khriplovich1997Book}
I.~Khriplovich \& S.~Lamoreaux.
\newblock \emph{CP violation without strangeness : electric dipole moments of
  particles, atoms, and molecules}.
\newblock Springer-Verlag, Berlin ; New York (1997).

\bibitem{Peccei1977}
R.~D. Peccei \& H.~R. Quinn.
\newblock Phys. Rev. Lett. \textbf{38} (1977) 1440.
\newblock \href {http://dx.doi.org/10.1103/PhysRevLett.38.1440}
  {\path{doi:10.1103/PhysRevLett.38.1440}}.

\bibitem{Wilczek1978}
F.~Wilczek.
\newblock Phys. Rev. Lett. \textbf{40} (1978) 279.
\newblock \href {http://dx.doi.org/10.1103/PhysRevLett.40.279}
  {\path{doi:10.1103/PhysRevLett.40.279}}.

\bibitem{Weinberg1978}
S.~Weinberg.
\newblock Phys. Rev. Lett. \textbf{40} (1978) 223.
\newblock \href {http://dx.doi.org/10.1103/PhysRevLett.40.223}
  {\path{doi:10.1103/PhysRevLett.40.223}}.

\bibitem{Baer2015}
H.~Baer, K.-Y. Choi, J.~Kim, \& L.~Roszkowski.
\newblock Phys. Rep. \textbf{555} (2015) 1.
\newblock \href {http://dx.doi.org/10.1016/j.physrep.2014.10.002}
  {\path{doi:10.1016/j.physrep.2014.10.002}}.

\bibitem{Jaeckel2012JHEP}
J.~Jaeckel \& V.~V. Khoze.
\newblock \jhep \textbf{2012} (2012) 1.
\newblock \href {http://dx.doi.org/10.1007/JHEP11(2012)115}
  {\path{doi:10.1007/JHEP11(2012)115}}.

\bibitem{Abel2006JHEP}
S.~Abel \& O.~Lebedev.
\newblock \jhep \textbf{01} (2006) 133.
\newblock \href {http://arxiv.org/abs/hep-ph/0508135}
  {\path{arXiv:hep-ph/0508135}}, \href
  {http://dx.doi.org/10.1088/1126-6708/2006/01/133}
  {\path{doi:10.1088/1126-6708/2006/01/133}}.

\bibitem{Ramsey1950PR}
N.~F. Ramsey.
\newblock Phys. Rev. \textbf{78} (1950) 695.
\newblock \href {http://dx.doi.org/10.1103/PhysRev.78.695}
  {\path{doi:10.1103/PhysRev.78.695}}.

\bibitem{Piegsa2009NIMA}
F.~Piegsa, B.~van~den Brandt, P.~Hautle, \& J.~Konter.
\newblock \nima \textbf{605} (2009) 5.
\newblock \href {http://dx.doi.org/10.1016/j.nima.2009.01.161}
  {\path{doi:10.1016/j.nima.2009.01.161}}.

\bibitem{Afach2015PRD}
S.~{Afach}, N.~J. {Ayres}, C.~A. {Baker}, et~al.
\newblock Phys. Rev. D \textbf{92} 052008.
\newblock \href {http://arxiv.org/abs/1506.06563} {\path{arXiv:1506.06563}},
  \href {http://dx.doi.org/10.1103/PhysRevD.92.052008}
  {\path{doi:10.1103/PhysRevD.92.052008}}.

\bibitem{Slichter1990book}
C.~P. Slichter.
\newblock \emph{Principles of Magnetic Resonance}, volume~1 of \emph{Springer
  Series in Solid-State Sciences}.
\newblock Springer-Verlag Berlin Heidelberg (1990).
\newblock \href {http://dx.doi.org/10.1007/978-3-662-09441-9}
  {\path{doi:10.1007/978-3-662-09441-9}}.

\bibitem{Harris2007arXiv}
P.~G. {Harris}.
\newblock ArXiv e-prints \href {http://arxiv.org/abs/0709.3100}
  {\path{arXiv:0709.3100}}.

\bibitem{Baker2011}
C.~Baker, G.~Ban, K.~Bodek, et~al.
\newblock Phys. Proc. \textbf{17} (2011) 159 \href{http://dx.doi.org/10.1016/j.phpro.2011.06.032}{\path{doi:10.1016/j.phpro.2011.06.032}}.

\bibitem{Golub1991}
R.~Golub, D.~Richardson, \& S.~Lamoreaux.
\newblock \emph{Ultra-Cold Neutrons}.
\newblock Adam Hilger, Bristol, Philadelphia, and New York (1991).

\bibitem{Ignatovich1990}
V.~Ignatovich.
\newblock \emph{The Physics of Ultracold Neutrons}.
\newblock Clarendon, Oxford (1990).

\bibitem{Steyerl1975NIM}
A.~{Steyerl}.
\newblock Nucl. Instr. Methods \textbf{125} (1975) 461.
\newblock \href {http://dx.doi.org/10.1016/0029-554X(75)90265-7}
  {\path{doi:10.1016/0029-554X(75)90265-7}}.

\bibitem{Steyerl1986PLA}
A.~{Steyerl}, H.~{Nagel}, F.-X. {Schreiber}, et~al.
\newblock Phys. Lett. A \textbf{116} (1986) 347.
\newblock \href {http://dx.doi.org/10.1016/0375-9601(86)90587-6}
  {\path{doi:10.1016/0375-9601(86)90587-6}}.

\bibitem{Trinks2000NIMA}
U.~{Trinks}, F.~J. {Hartmann}, S.~{Paul}, \& W.~{Schott}.
\newblock \nima \textbf{440} (2000) 666.
\newblock \href {http://dx.doi.org/10.1016/S0168-9002(99)01059-1}
  {\path{doi:10.1016/S0168-9002(99)01059-1}}.

\bibitem{Saunders2013RSI}
A.~Saunders, M.~Makela, Y.~Bagdasarova, et~al.
\newblock \rsi \textbf{84} (2013) 013304.
\newblock \href {http://dx.doi.org/10.1063/1.4770063}
  {\path{doi:10.1063/1.4770063}}.

\bibitem{Lauss2014}
B.~{Lauss}.
\newblock Physics Procedia \textbf{51} (2014) 98.
\newblock \href {http://dx.doi.org/10.1016/j.phpro.2013.12.022}
  {\path{doi:10.1016/j.phpro.2013.12.022}}.

\bibitem{Piegsa2014}
F.~M. Piegsa, M.~Fertl, S.~N. Ivanov, et~al.
\newblock \prc \textbf{90} (2014) 015501.
\newblock \href {http://dx.doi.org/10.1103/PhysRevC.90.015501}
  {\path{doi:10.1103/PhysRevC.90.015501}}.

\bibitem{Serebrov2014TPL}
A.~P. {Serebrov}, A.~K. {Fomin}, M.~S. {Onegin}, et~al.
\newblock Tech. Phys. Lett. \textbf{40} (2014) 10.
\newblock \href {http://dx.doi.org/10.1134/S1063785014010118}
  {\path{doi:10.1134/S1063785014010118}}.

\bibitem{Masuda2015JPS}
Y.~Masuda, K.~Hatanaka, S.-C. Jeong, et~al.
\newblock JPS Conf. Proc. \textbf{8} (2015) 026002.
\newblock \href {http://dx.doi.org/10.7566/JPSCP.8.026002}
  {\path{doi:10.7566/JPSCP.8.026002}}.

\bibitem{Zimmer2015PRC}
O.~Zimmer \& R.~Golub.
\newblock \prc \textbf{92} (2015) 015501.
\newblock \href {http://dx.doi.org/10.1103/PhysRevC.92.015501}
  {\path{doi:10.1103/PhysRevC.92.015501}}.

\bibitem{Golub1975}
R.~{Golub} \& J.~M. {Pendlebury}.
\newblock \pla \textbf{53} (1975) 133.
\newblock \href {http://dx.doi.org/10.1016/0375-9601(75)90500-9}
  {\path{doi:10.1016/0375-9601(75)90500-9}}.

\bibitem{Golub1977}
R.~{Golub} \& J.~M. {Pendlebury}.
\newblock \pla \textbf{62} (1977) 337.
\newblock \href {http://dx.doi.org/10.1016/0375-9601(77)90434-0}
  {\path{doi:10.1016/0375-9601(77)90434-0}}.

\bibitem{Kirch2010}
K.~Kirch, B.~Lauss, P.~Schmidt-Wellenburg, \& G.~Zsigmond.
\newblock Nucl. Phys. News \textbf{{20:1}} (2010) 17.
\newblock \href {http://dx.doi.org/10.1080/10619121003626724}
  {\path{doi:10.1080/10619121003626724}}.

\bibitem{Korobkina2002}
E.~{Korobkina}, R.~{Golub}, B.~W. {Wehring}, \& A.~R. {Young}.
\newblock \pla \textbf{301} (2002) 462.
\newblock \href {http://arxiv.org/abs/cond-mat/0204635}
  {\path{arXiv:cond-mat/0204635}}, \href
  {http://dx.doi.org/10.1016/S0375-9601(02)01052-6}
  {\path{doi:10.1016/S0375-9601(02)01052-6}}.

\bibitem{Schmidt-Wellenburg2009}
P.~{Schmidt-Wellenburg}, K.~H. {Andersen}, \& O.~{Zimmer}.
\newblock \nima \textbf{611} (2009) 259.
\newblock \href {http://arxiv.org/abs/0811.4332} {\path{arXiv:0811.4332}},
  \href {http://dx.doi.org/10.1016/j.nima.2009.07.085}
  {\path{doi:10.1016/j.nima.2009.07.085}}.

\bibitem{Atchison2007PRL}
F.~{Atchison}, B.~{Blau}, K.~{Bodek}, et~al.
\newblock \prl \textbf{99} 262502.
\newblock \href {http://dx.doi.org/10.1103/PhysRevLett.99.262502}
  {\path{doi:10.1103/PhysRevLett.99.262502}}.

\bibitem{Kirch2014FPUA}
K.~Kirch \& P.~Schmidt-Wellenburg.
\newblock Procedings FPUA
  \urlprefix\url{http://xqw.hep.okayama-u.ac.jp/kakenhi/files/4813/9814/9853/11-Kirch.pdf}.

\bibitem{Luschikov1984NIMA}
V.~I. {Luschikov} \& Y.~V. {Taran}.
\newblock \nima \textbf{228} (1984) 159.
\newblock \href {http://dx.doi.org/10.1016/0168-9002(84)90025-1}
  {\path{doi:10.1016/0168-9002(84)90025-1}}.

\bibitem{Geltenbort2009NIMA}
P.~{Geltenbort}, L.~{G{\"o}ltl}, R.~{Henneck}, et~al.
\newblock \nima \textbf{608} (2009) 132.
\newblock \href {http://dx.doi.org/10.1016/j.nima.2009.06.038}
  {\path{doi:10.1016/j.nima.2009.06.038}}.

\bibitem{Baker2014}
C.~Baker, Y.~Chibane, M.~Chouder, et~al.
\newblock \nima \textbf{736} (2014) 184.

\bibitem{Serebrov2015PRC}
A.~P. {Serebrov}, E.~A. {Kolomenskiy}, A.~N. {Pirozhkov}, et~al.
\newblock \prc \textbf{92} 055501.
\newblock \href {http://dx.doi.org/10.1103/PhysRevC.92.055501}
  {\path{doi:10.1103/PhysRevC.92.055501}}.

\bibitem{Afach2015EPJA}
S.~{Afach}, G.~{Ban}, G.~{Bison}, et~al.
\newblock \epja \textbf{51} (2015) 143.
\newblock \href {http://arxiv.org/abs/1502.06876} {\path{arXiv:1502.06876}},
  \href {http://dx.doi.org/10.1140/epja/i2015-15143-7}
  {\path{doi:10.1140/epja/i2015-15143-7}}.

\bibitem{NeutronDataBooklet}
A.-J. Dianoux \& G.~Lander (editors).
\newblock \emph{Neutron Data Booklet}.
\newblock OCP Science (2003).

\bibitem{Morris2009NIMA}
C.~L. {Morris}, T.~J. {Bowles}, J.~{Gonzales}, et~al.
\newblock \nima \textbf{599} (2009) 248.
\newblock \href {http://dx.doi.org/10.1016/j.nima.2008.11.099}
  {\path{doi:10.1016/j.nima.2008.11.099}}.

\bibitem{Klein2011NIMA}
M.~{Klein} \& C.~J. {Schmidt}.
\newblock \nima \textbf{628} (2011) 9.
\newblock \href {http://dx.doi.org/10.1016/j.nima.2010.06.278}
  {\path{doi:10.1016/j.nima.2010.06.278}}.

\bibitem{Salvat2012NIMA}
D.~J. {Salvat}, C.~L. {Morris}, Z.~{Wang}, et~al.
\newblock \nima \textbf{691} (2012) 109.
\newblock \href {http://dx.doi.org/10.1016/j.nima.2012.06.041}
  {\path{doi:10.1016/j.nima.2012.06.041}}.

\bibitem{Baker2003NIMA}
C.~A. {Baker}, S.~N. {Balashov}, K.~{Green}, et~al.
\newblock \nima \textbf{501} (2003) 517.
\newblock \href {http://dx.doi.org/10.1016/S0168-9002(03)00384-X}
  {\path{doi:10.1016/S0168-9002(03)00384-X}}.

\bibitem{Lasakov2005NIMA}
M.~S. {Lasakov}, A.~P. {Serebrov}, A.~K. {Khusainov}, et~al.
\newblock \nima \textbf{545} (2005) 301.
\newblock \href {http://dx.doi.org/10.1016/j.nima.2005.01.317}
  {\path{doi:10.1016/j.nima.2005.01.317}}.

\bibitem{Lauer2011EPJA}
T.~{Lauer}, P.~{Geltenbort}, P.~{Hoebel}, et~al.
\newblock \epja \textbf{47} 150.
\newblock \href {http://dx.doi.org/10.1140/epja/i2011-11150-0}
  {\path{doi:10.1140/epja/i2011-11150-0}}.

\bibitem{Ban2009NIMA}
G.~{Ban}, K.~{Bodek}, T.~{Lefort}, et~al.
\newblock \nima \textbf{611} (2009) 280.
\newblock \href {http://dx.doi.org/10.1016/j.nima.2009.07.083}
  {\path{doi:10.1016/j.nima.2009.07.083}}.

\bibitem{Goeltl2013EPJA}
L.~{G{\"o}ltl}, Z.~{Chowdhuri}, M.~{Fertl}, et~al.
\newblock \epja \textbf{49} 9.
\newblock \href {http://dx.doi.org/10.1140/epja/i2013-13009-8}
  {\path{doi:10.1140/epja/i2013-13009-8}}.

\bibitem{Jamieson2015NIMA}
B.~{Jamieson} \& L.~A. {Rebenitsch}.
\newblock \nima \textbf{790} (2015) 6.
\newblock \href {http://arxiv.org/abs/1502.01392} {\path{arXiv:1502.01392}},
  \href {http://dx.doi.org/10.1016/j.nima.2015.04.022}
  {\path{doi:10.1016/j.nima.2015.04.022}}.

\bibitem{Ban2016EPJA}
G.~{Ban}, G.~{Bison}, K.~{Bodek}, et~al.
\newblock ArXiv e-prints \href {http://arxiv.org/abs/1606.07432}
  {\path{arXiv:1606.07432}}.

\bibitem{Wang2015NIMA}
Z.~{Wang}, M.~A. {Hoffbauer}, C.~L. {Morris}, et~al.
\newblock \nima \textbf{798} (2015) 30.
\newblock \href {http://arxiv.org/abs/1503.03424} {\path{arXiv:1503.03424}},
  \href {http://dx.doi.org/10.1016/j.nima.2015.07.010}
  {\path{doi:10.1016/j.nima.2015.07.010}}.

\bibitem{Ito2014arXiv}
T.~M. {Ito}, D.~H. {Beck}, S.~M. {Clayton}, et~al.
\newblock ArXiv e-prints \href {http://arxiv.org/abs/1401.5435}
  {\path{arXiv:1401.5435}}.

\bibitem{Leung2016Mainz}
K.~K.~H. Leung, W.~C. Griffith, \& Z.~Tang.
\newblock Mainz Workshop: Probing Fundamental Symmetries and Interactions with
  UCN
  \urlprefix\url{https://indico.mitp.uni-mainz.de/event/59/timetable/#20160413.detailed}.

\bibitem{Abele2006}
H.~{Abele}, D.~{Dubbers}, H.~{H{\"a}se}, et~al.
\newblock \nima \textbf{562} (2006) 407.
\newblock \href {http://arxiv.org/abs/arXiv:nucl-ex/0510072}
  {\path{arXiv:arXiv:nucl-ex/0510072}}, \href
  {http://dx.doi.org/10.1016/j.nima.2006.03.020}
  {\path{doi:10.1016/j.nima.2006.03.020}}.

\bibitem{Baker2010JPhCS}
C.~A. {Baker}, S.~N. {Balashov}, V.~{Francis}, et~al.
\newblock J. Phys. Conf. Ser. \textbf{251} 012055.
\newblock \href {http://dx.doi.org/10.1088/1742-6596/251/1/012055}
  {\path{doi:10.1088/1742-6596/251/1/012055}}.

\bibitem{Golub1994PhR}
R.~{Golub} \& S.~K. {Lamoreaux}.
\newblock \physrep \textbf{237} (1994) 1.
\newblock \href {http://dx.doi.org/10.1016/0370-1573(94)90084-1}
  {\path{doi:10.1016/0370-1573(94)90084-1}}.

\bibitem{Esler2007PRC}
A.~{Esler}, J.~C. {Peng}, D.~{Chandler}, et~al.
\newblock \prc \textbf{76} 051302.
\newblock \href {http://arxiv.org/abs/nucl-ex/0703029}
  {\path{arXiv:nucl-ex/0703029}}, \href
  {http://dx.doi.org/10.1103/PhysRevC.76.051302}
  {\path{doi:10.1103/PhysRevC.76.051302}}.

\bibitem{Afach2014JAP}
S.~Afach et~al.
\newblock J. Appl. Phys. \textbf{116} (2014) 084510.
\newblock \href {http://arxiv.org/abs/1408.6752} {\path{arXiv:1408.6752}},
  \href {http://dx.doi.org/10.1063/1.4894158} {\path{doi:10.1063/1.4894158}}.

\bibitem{Sumner1987JPhD}
T.~J. {Sumner}, J.~M. {Pendlebury}, \& K.~F. {Smith}.
\newblock J. Phys. D: Appl. Phys. \textbf{20} (1987) 1095.
\newblock \href {http://dx.doi.org/10.1088/0022-3727/20/9/001}
  {\path{doi:10.1088/0022-3727/20/9/001}}.

\bibitem{Altarev2014RSI}
I.~{Altarev}, E.~{Babcock}, D.~{Beck}, et~al.
\newblock Rev. Sci. Instrum. \textbf{85} (2014) 075106.
\newblock \href {http://arxiv.org/abs/1403.6467} {\path{arXiv:1403.6467}},
  \href {http://dx.doi.org/10.1063/1.4886146} {\path{doi:10.1063/1.4886146}}.

\bibitem{Afach2015OExpress}
S.~{Afach}, G.~{Ban}, G.~{Bison}, et~al.
\newblock Optics Express \textbf{23} (2015) 22108.
\newblock \href {http://arxiv.org/abs/1507.08523} {\path{arXiv:1507.08523}},
  \href {http://dx.doi.org/10.1364/OE.23.022108}
  {\path{doi:10.1364/OE.23.022108}}.

\bibitem{Budker2007}
D.~{Budker} \& M.~{Romalis}.
\newblock Nature Physics \textbf{3} (2007) 227.
\newblock \href {http://arxiv.org/abs/physics/0611246}
  {\path{arXiv:physics/0611246}}, \href {http://dx.doi.org/10.1038/nphys566}
  {\path{doi:10.1038/nphys566}}.

\bibitem{Burghoff2007}
M.~Burghoff, S.~Hartwig, W.~Kilian, et~al.
\newblock IEEE Trans. App. Supercond. \textbf{17} (2007) 846.
\newblock \href {http://dx.doi.org/10.1109/TASC.2007.898203}
  {\path{doi:10.1109/TASC.2007.898203}}.

\bibitem{Smith1990PLB}
K.~F. {Smith}, N.~{Crampin}, J.~M. {Pendlebury}, et~al.
\newblock \plb \textbf{234} (1990) 191.
\newblock \href {http://dx.doi.org/10.1016/0370-2693(90)92027-G}
  {\path{doi:10.1016/0370-2693(90)92027-G}}.

\bibitem{Altarev1996}
I.~Altarev, Y.~Borisov, N.~Borovikova, et~al.
\newblock Phys. Atom. Nucl. \textbf{59} (1996) 1152.
\newblock \urlprefix\url{http://adsabs.harvard.edu/abs/1996PAN....59.1152A}.

\bibitem{Groeger2006}
S.~{Groeger}, G.~{Bison}, J.~{Schenker}, et~al.
\newblock \epjd \textbf{38} (2006) 239.
\newblock \href {http://dx.doi.org/10.1140/epjd/e2006-00037-y}
  {\path{doi:10.1140/epjd/e2006-00037-y}}.

\bibitem{Afach2014PLB}
S.~Afach, C.~Baker, G.~Ban, et~al.
\newblock \plb \textbf{739} (2014) 128 .
\newblock \href
  {http://dx.doi.org/http://dx.doi.org/10.1016/j.physletb.2014.10.046}
  {\path{doi:http://dx.doi.org/10.1016/j.physletb.2014.10.046}}.

\bibitem{Lamoreaux2009}
S.~K. {Lamoreaux} \& R.~{Golub}.
\newblock J. Phys. G: Nucl. Phys. \textbf{36} 104002.
\newblock \href {http://dx.doi.org/10.1088/0954-3899/36/10/104002}
  {\path{doi:10.1088/0954-3899/36/10/104002}}.

\bibitem{Golub1983JPL}
R.~Golub.
\newblock J. Physique Lett. \textbf{44} (1983) 321.
\newblock \href {http://dx.doi.org/10.1051/jphyslet:01983004409032100}
  {\path{doi:10.1051/jphyslet:01983004409032100}}.

\bibitem{Ramsey1984}
N.~F. Ramsey.
\newblock Acta Physica Hungarica \textbf{55} (1984) 117.
\newblock \href {http://dx.doi.org/10.1007/BF03155926}
  {\path{doi:10.1007/BF03155926}}.

\bibitem{Green1998}
K.~{Green}, P.~G. {Harris}, P.~{Iaydjiev}, et~al.
\newblock \nima \textbf{404} (1998) 381.
\newblock \href {http://dx.doi.org/10.1016/S0168-9002(97)01121-2}
  {\path{doi:10.1016/S0168-9002(97)01121-2}}.

\bibitem{Afach2015PRL}
S.~{Afach}, N.~J. {Ayres}, G.~{Ban}, et~al.
\newblock \prl \textbf{115} 162502.
\newblock \href {http://arxiv.org/abs/1506.00446} {\path{arXiv:1506.00446}},
  \href {http://dx.doi.org/10.1103/PhysRevLett.115.162502}
  {\path{doi:10.1103/PhysRevLett.115.162502}}.

\bibitem{Rabi1954}
I.~Rabi, N.~Ramsey, \& J.~Schwinger.
\newblock \rmp \textbf{26} (1954) 167.
\newblock \href {http://dx.doi.org/10.1103/RevModPhys.26.167}
  {\path{doi:10.1103/RevModPhys.26.167}}.

\bibitem{Zenner2013}
J.~{Zenner}.
\newblock \emph{The search for the neutron electric dipole moment}.
\newblock Ph.D. thesis, Johannes Gutenberg Universit\"at, Mainz (2013).
\newblock \urlprefix\url{http://d-nb.info/1041244924/34}.

\bibitem{Piegsa2013PRC}
F.~M. {Piegsa}.
\newblock \prc \textbf{88} 045502.
\newblock \href {http://arxiv.org/abs/1309.1959} {\path{arXiv:1309.1959}},
  \href {http://dx.doi.org/10.1103/PhysRevC.88.045502}
  {\path{doi:10.1103/PhysRevC.88.045502}}.

\bibitem{Pignol2012PRA}
G.~{Pignol} \& S.~{Roccia}.
\newblock \pra \textbf{85} 042105.
\newblock \href {http://arxiv.org/abs/1201.0699} {\path{arXiv:1201.0699}},
  \href {http://dx.doi.org/10.1103/PhysRevA.85.042105}
  {\path{doi:10.1103/PhysRevA.85.042105}}.

\bibitem{Pendlebury2004}
J.~M. Pendlebury, W.~Heil, Y.~Sobolev, et~al.
\newblock \pra \textbf{70} (2004) 032102.
\newblock \href {http://dx.doi.org/10.1103/PhysRevA.70.032102}
  {\path{doi:10.1103/PhysRevA.70.032102}}.

\bibitem{Lamoreaux2005}
S.~K. {Lamoreaux} \& R.~{Golub}.
\newblock \pra \textbf{71} (2005) 032104.
\newblock \href {http://arxiv.org/abs/nucl-ex/0407005}
  {\path{arXiv:nucl-ex/0407005}}, \href
  {http://dx.doi.org/10.1103/PhysRevA.71.032104}
  {\path{doi:10.1103/PhysRevA.71.032104}}.

\bibitem{Barabanov2006}
A.~L. {Barabanov}, R.~{Golub}, \& S.~K. {Lamoreaux}.
\newblock \pra \textbf{74} (2006) 052115.
\newblock \href {http://dx.doi.org/10.1103/PhysRevA.74.052115}
  {\path{doi:10.1103/PhysRevA.74.052115}}.

\bibitem{Afach2015EPJD}
S.~{Afach}, C.~A. {Baker}, G.~{Ban}, et~al.
\newblock \epjd \textbf{69} 225.
\newblock \href {http://arxiv.org/abs/1503.08651} {\path{arXiv:1503.08651}},
  \href {http://dx.doi.org/10.1140/epjd/e2015-60207-4}
  {\path{doi:10.1140/epjd/e2015-60207-4}}.

\bibitem{Berry1984}
M.~V. Berry.
\newblock Proc. Roy. Soc. Lond. \textbf{A392} (1984) 45.
\newblock \href {http://dx.doi.org/10.1098/rspa.1984.0023}
  {\path{doi:10.1098/rspa.1984.0023}}.

\bibitem{Commins1991}
E.~D. {Commins}.
\newblock \ajp \textbf{59} (1991) 1077.
\newblock \href {http://dx.doi.org/10.1119/1.16616}
  {\path{doi:10.1119/1.16616}}.

\bibitem{Masuda2012PLA}
Y.~{Masuda}, K.~{Asahi}, K.~{Hatanaka}, et~al.
\newblock \pla \textbf{376} (2012) 1347.
\newblock \href {http://dx.doi.org/10.1016/j.physleta.2012.02.056}
  {\path{doi:10.1016/j.physleta.2012.02.056}}.

\bibitem{Fedorov2011PhyB}
V.~V. {Fedorov}, V.~V. {Voronin}, \& Y.~P. {Braginetz}.
\newblock Physica B \textbf{406} (2011) 2370.
\newblock \href {http://dx.doi.org/10.1016/j.physb.2010.10.080}
  {\path{doi:10.1016/j.physb.2010.10.080}}.

\bibitem{Ageron1978}
P.~{Ageron}, W.~{Mampe}, R.~{Golub}, \& J.~M. {Pendelbury}.
\newblock \pla \textbf{66} (1978) 469.
\newblock \href {http://dx.doi.org/10.1016/0375-9601(78)90399-7}
  {\path{doi:10.1016/0375-9601(78)90399-7}}.

\bibitem{Altarev2012a}
I.~Altarev, D.~Beck, S.~Chesnevskaya, et~al.
\newblock Nuovo Cim. \textbf{C035N04} (2012) 122.
\newblock \href {http://dx.doi.org/10.1393/ncc/i2012-11271-0}
  {\path{doi:10.1393/ncc/i2012-11271-0}}.

\bibitem{Kolarkar2010AIP}
A.~Kolarkar.
\newblock AIP Conf. Proc. \textbf{1200} (2010) 861.
\newblock \href {http://dx.doi.org/10.1063/1.3327748}
  {\path{doi:10.1063/1.3327748}}.

\bibitem{Matsuta2013AIP}
K.~Matsuta et~al.
\newblock AIP Conf. Proc. \textbf{1560} (2013) 152.
\newblock \href {http://dx.doi.org/10.1063/1.4826742}
  {\path{doi:10.1063/1.4826742}}.

\end{thebibliography}

\end{document}